\documentclass[a4paper, 11pt]{article}
\usepackage[a4paper,hmargin={2.5cm,2.5cm},vmargin={3cm,3cm}]{geometry}

\usepackage{amsmath,amssymb,graphicx}
\usepackage{amsthm}

\renewcommand{\Re}{\mathrm{Re}}
\renewcommand{\Im}{\mathrm{Im}}

\newtheorem{conjecture}{Conjecture}

\title{A numerical study of blow-up and stability of line solitons for the Novikov-Veselov equation}
\date{}
\author{Anna Kazeykina\footnote{Laboratoire de Math\'ematiques d'{}Orsay,
Universit\'e Paris-Sud. E-mail: anna.kazeykina at math.u-psud.fr} \quad and \quad  Christian Klein\footnote{Institut de Math\'ematiques de Bourgogne, Universit\'e de Bourgogne. E-mail: christian.klein at u-bourgogne.fr}}
\begin{document}

\maketitle
\begin{abstract}
We study numerically the evolution of perturbed Korteweg-de Vries 
solitons and of well localized initial data by the Novikov-Veselov 
(NV) equation at different levels of the ``energy'' parameter $ E $. 
We show that as $ |E| \to \infty $, NV behaves, as expected, 
similarly to its formal limit, the Kadomtsev-Petviashvili equation. 
However at intermediate regimes, i.e. when $ | E | $ is not very 
large, more varied scenarios are possible, in particular, 
\emph{blow-ups} are observed. The mechanism of the blow-up is studied. 
\end{abstract}

\section{Introduction}
In the present paper we study numerically well-posedeness and stability issues for the Novikov-Veselov (NV) equation:
\begin{subequations}
\label{NV}
\begin{align}
\label{NV_eq}
& \partial_t v = 4 \Re ( 4 \partial_z^3 v + \partial_z( v w ) - E 
\partial_{ z } w ),\\
\label{w_def}
& \partial_{ \bar z } w = - 3 \partial_z v, \quad v = \bar v, \text{ i.e. } v \text{ is a real-valued function, } \quad E \in \mathbb{R},\\
\label{misc}
& v = v( x, y, t ), \quad w = w( x, y, t ),  \quad z=x+iy, \quad ( x, y ) \in \mathbb{R}^2, \quad t \in \mathbb{R},
\end{align}
\end{subequations}
where
\begin{equation}
\label{derivatives}
\partial_t = \frac{ \partial }{ \partial t }, \quad \partial_z = \frac{ 1 }{ 2 } \left( \frac{ \partial }{ \partial x } - i \frac{ \partial }{ \partial y } \right), \quad \partial_{ \bar z } = \frac{ 1 }{ 2 } \left( \frac{ \partial }{ \partial x } + i \frac{ \partial }{ \partial y } \right).
\end{equation}

In terms of the variables $ x, y $, equation (\ref{w_def}) reads
\begin{equation}
\label{w_xy}
\partial_y w_1+\partial_x w_2 =3~\partial_y v, \quad \partial_y w_2- \partial_x w_1 =3~ \partial_x v,
\end{equation}
where $ w_1 = \Re w $, $ w_2 = \Im w $. If we identify the complex-valued function $ w = w_1 + i w_2 $ with a vector-valued function $ w = ( w_1, w_2 ) $, then (\ref{NV_eq}) is written:
\begin{equation}
\label{NV_xy}
\partial_t v = 2\left[ \partial_x (\partial_x^2 v-3 \partial_y^2v) +\nabla .(vw) -E\nabla .w \right].
\end{equation}
The real-valued parameter $ E $ is referred to as \emph{energy} for the reasons explained below.

Note that (\ref{NV}) can be viewed as a nonlocal equation in $ v $, with the nonlocal term given by $ w = - 3 \partial_{ \bar z }^{ -1 } \partial_z v $. The operator $ \partial_{ \bar z }^{ -1 } $ can be defined via the Fourier transform by using (\ref{w_xy}):
\begin{equation}
\label{fourier_w}
\hat w_1( \xi_1, \xi_2 ) = \frac{ 3 ( \xi_2^2 - \xi_1^2 ) }{ \xi_1^2 
+ \xi_2^2 }\hat v( \xi_1, \xi_2 ), \quad \hat w_2( \xi_1, \xi_2 ) = \frac{ 6 \xi_1 \xi_2 }{ \xi_1^2 + \xi_2^2 } \hat v( \xi_1, \xi_2 ),
\end{equation}
where $\hat{w}_{i}$, $i=1,2$ (or $\mathcal{F}w_{i}$) denotes the two-dimensional Fourier 
transform of $w_{i}$, and where $(\xi_{1},\xi_{2})$ are the dual 
variables of $(x,y)$.

Equation (\ref{NV}) is a $ ( 2 + 1 ) $-dimensional analog of the 
renowned Korteweg-de Vries (KdV) equation (note that if $ v( x, y, t ) = v( x, t ) $, $ w( x, y, t ) = w( x, t ) $, then (\ref{NV}) is reduced to the classic KdV equation). Equation (\ref{NV}) has been derived as a nonlinear PDE integrable via the inverse scattering transform (IST) for the two--dimensional Schr\"odinger equation at fixed energy $ E $  
\begin{equation}
\label{schrodinger}
 L \psi = E \psi, \quad L = - \Delta + v( x, y ), \quad E = E_{ fixed }
\end{equation}
(see \cite{M}, \cite{NV1}, \cite{NV2}). It was shown, in particular, that for the Schr\"odinger operator $ L $ from (\ref{schrodinger}) there exist operators $ A $, $ B $ (Manakov L--A--B triple) such that (\ref{NV}) is equivalent to
\begin{equation*}
 \frac{ \partial ( L - E ) }{ \partial t } = [ L - E, A ] + B ( L - E ),
\end{equation*}
where $ [ \cdot, \cdot ] $ is the commutator. More precisely, operators $ A $ and $ B $ are given by the following formulas:
\begin{equation}
\label{int_A-B}
\begin{array}{l}
A = - 8 \partial_{ z }^3 - 2 w \partial_{ z } - 8 \partial_{ \bar z }^3 - 2 \bar w \partial_{ \bar z }, \\
B = 2 \partial_{ \bar z } w + 2 \partial_{ \bar z } \bar w,
\end{array} \text{ where } w \text{ is defined via (\ref{misc})}.
\end{equation}
For a detailed description of the IST method for NV the reader is 
referred to the review paper \cite{G}.

There has not been so far any physical derivation of the NV equation 
(\ref{NV}). Note, however, that the dispersionless NV at $ E = 0 $ was derived in a model of nonlinear geometrical optics (see \cite{KM3}). Also it was shown that the stationary NV at $ E = 0 $ describes isothermally asymptotic surfaces in projective geometry (see \cite{F}).

Though the NV equation itself lacks a physical interpretation, the 
more physically relevant Kadomtsev-Petviashvili (KP) equation can be 
regarded as a high energy limit of the Novikov-Veselov equation. More 
precisely, in equation (\ref{NV}) put  $ E = \pm \kappa^2 $, dilate 
the second variable: $ y = \kappa Y $,   and take for $ w $ the following ansatz:
\begin{equation*}
w = \mp 3 \kappa^2 - 3 v + 6 i \partial_x^{-1} \partial_Y \nu \frac{ 1 }{ \kappa } + 6 ( \partial_x^{-1} )^2 \partial_Y^2 \nu \frac{ 1 }{ \kappa^2 } 
\end{equation*}
(note that with this ansatz the equation (\ref{w_def}) is satisfied up to terms of order greater than $ 1/\kappa^2 $).
Then, in the limit $ \kappa \to \infty $, the function $ \nu( x, Y, t ) = v( x, \kappa Y, t ) $ satisfies the following equation:
\begin{equation*}
\label{KP}
 \partial_{t} \nu = 2 \partial_{xxx} \nu - 12 v \partial_{x} \nu \mp 24 \partial_{x}^{-1} \partial_{Y}^2 \nu
\end{equation*}
and the function
\begin{equation}
\label{uv_relation}
u( x, y, t ) = - \nu( -x, 2 y, \frac{ 1 }{ 2 } t )
\end{equation} is a solution of KP in its classical form
\begin{equation}
\label{KP_classic}
\partial_t u + 6 u \partial_x u + \partial_{ xxx } u = \pm 3 \partial_{ x }^{ -1 } \partial_{ y }^2 u.
\end{equation}
Thus we see that, at high energy $ E $ limit, after an appropriate 
dilation of the $ y $ variable, the NV equation at positive energy 
becomes KPI ($+$ sign on the right hand sign of (\ref{KP_classic})), 
and the NV equation at negative energy becomes KPII ($-$ sign on the right hand sign of (\ref{KP_classic})) (see \cite{KM} for details).

It was also shown in \cite{G} that as $ E \to +\infty $, the standard 
scattering data of the two-dimensional stationary Schr\"odinger 
equation (\ref{schrodinger}) converge to the scattering data for the 
time-dependent one-dimensional Schr\"odinger equation (arising in the IST method for KPI) and as $ E \to -\infty $, they converge to the scattering data for the heat equation (arising in the IST method for KPII).

From this perspective, NV can be viewed as an intermediate model 
between two regimes: the semilinear regime of KPII and the 
quasilinear regime of KPI (see \cite{KS1,KS2} and references therein for 
more information on KP). We expect that at high values of the energy 
$ E $ the behaviour of solutions for NV is reminiscent of that for 
KP. This is indeed confirmed by the numerical study performed in the 
present paper. At intermediate values of the energy $ E $, however, the situation is different. We demonstrate that a richer variety of possible behaviours occur when the values of $ E $ are not too large.

The paper is organised as follows. In Section \ref{review_section} we 
review briefly some theoretical results on the NV equation, with a 
focus on the properties that will be investigated numerically in the 
present paper. In Section \ref{dyn_resc} we review the applied numerical 
techniques and introduce the dynamic rescaling of the NV equation 
used to study the blow-up phenomena. In Section \ref{soliton_stab} we 
study numerically the stability of a KdV soliton with respect to the 
NV equation. In Section \ref{localized} we study the evolution of the 
localized initial data under the NV dynamics. In Section 
\ref{sec_conc} we summarize the main results and add some concluding remarks.

\section{Review of theoretical and numerical results}
\label{review_section}
\subsection{Conservation laws and symmetries}
The Novikov-Veselov equation has an infinite number of conserved 
quantities. Here are the first three (each integral is taken over $\mathbb{C} \cong \mathbb{R}^2$): the $L^1$ integral
\begin{equation*}
\int v(t) dxdy =\int v(0)dxdy, \quad \hbox{(real-valued),}
\end{equation*}
the ``mass''
\begin{equation}
\label{mass}
M[v](t) :=\int vw(t) dxdy
 =\int vw(0)dxdy, \quad \hbox{(complex-valued, two identities),}
\end{equation}
and the ``energy'':
\begin{equation}
\label{H_energy}
H[v](t) := \int \Big[  6\,  \partial_z w \partial_z v  + E w^2 - vw^2  \Big] (t)dxdy \\
=H[v](0),\quad \hbox{(complex-valued, two identities).}
\end{equation}
The NV equation does not seem to have  conserved quantities of 
definite sign that control the long-time dynamics. Note that 
though in KdV reduction and KP limit, the conservation law given by 
(\ref{mass}) corresponds to the $ L^2 $ norm, in the NV case this 
quantity does not give any bound for the $ L^2 $ norm. Indeed, using representation (\ref{fourier_w}) it is easy to see that for spherically symmetric solutions $ v $ the quantity $ M $ is zero. Thus it cannot give any useful control on the $ L^2 $ norm of the solution.

Equation (\ref{NV}) possesses the following scaling symmetry: if $v(t,x,y)$ is solution of NV, then for all $\lambda>0$,
\begin{equation}\label{invariance}
v_\lambda(x,y,t) := \lambda^2 v(\lambda x,\lambda y,\lambda^3 t), \quad \hbox{and} \quad w_\lambda(x,y,t) := \lambda^2 w(\lambda x,\lambda y,\lambda^3 t)
\end{equation}
satisfy (\ref{NV}) with
\begin{equation*}
E_\lambda := E \lambda^2.
\end{equation*}
Note that the Sobolev space $ H^{-1}(\mathbb{R}^2) $ is scaling invariant for NV at $ E = 0 $.  

Equation (\ref{NV}) at $ E = 0 $ is invariant with respect to 
rotations by $ \frac{ 2 \pi }{ 3 } $ and $ \frac{ 4 \pi }{ 3 } $ (see \cite{Cetal}).

\subsection{Well-posedness of Cauchy problem, blow-up solutions}
The inverse scattering theory for the Novikov-Veselov equation implies that in the case $ E \neq 0 $ the Cauchy problem for the NV equation with sufficiently small, regular initial data possesses a global solution (see \cite{N, G}). For the case $ E = 0 $, an additional spectral assumption on the initial data has to be made in order to ensure the existence of global solutions to the NV equation: more precisely, one can show via the IST techniques that if the initial data $ v_0 $ satisfy $ - \Delta + v_0 \geq 0 $ and are sufficiently regular and localized, then the solution of the corresponding Cauchy problem exists globally in time (\cite{Na, P, MP}).

The first result on the well-posedness of the Cauchy problem for the 
NV equation without any spectral assumption or an assumption of 
smallness of initial data was obtained in \cite{A}. It was shown that 
the NV equation at $ E = 0 $ is well-posed in $ H^s $ with $ s > \frac{ 1 }{ 2 } $, locally in time. The equation is ill-posed in $ H^s $ if $ s < -1 $.
Note also that this work did not make use of the integrable nature of the equation. The techniques employed are dispersive estimates and Bourgain space method. 

The ideas of \cite{A} were developed in \cite{KM, KM2} to show the 
local (in time) well-posedness of the NV equation in the case $ E 
\neq 0 $ in $ H^s $, $ s > \frac{ 1 }{ 2 } $ (the results are based 
on the derivation of smoothing and dispersive estimates for a nonlocal two-dimensional symbol). The authors also obtain the dependence of the time of existence on the energy. It is shown that if $ s > \frac{ 7 }{ 8 } $, then the time of existence of a solution is proportional to $ | E |^{ \alpha } $, where $ \alpha $ is some positive real number.

Though the local existence theory for NV equation has been developed, there is little hope to obtain global results via the use of the conservation laws. 
For $ E = 0 $ an explicit construction confirms the fact that the $ L^2 $ norm cannot be bounded by conservation laws: for NV at $ E = 0 $ in \cite{KM2} the authors construct examples of solutions which blow up in $ L^2 $ norm in infinite time. Define 
\begin{equation}\label{hopper}
Q_{n}(z,t) := -8 \partial_z \partial_{\bar z} \log \big(1+ |P_n(z,t)|^2 \big) = -\frac{8|\partial_z P_n(z,t)|^2}{(1+|P_n(z,t)|^2)^2},
\end{equation}
where $ P_n( z,t ) $ are Gould-Hopper polynomials defined in terms of the complex-valued Airy symbol, for $\lambda\in \mathbb{R} $:
\[
e^{\lambda z + 8\lambda^3 t} =: \sum_{n\geq 0} P_n(z,t) \frac{\lambda^n}{n!}, \qquad P_n(z,0) =z^n.
\] 
In particular,
\begin{equation*}
P_0 =1, \quad P_1 = z, \quad P_2 = z^2, \quad
P_3 = z^3 + 48 t,\quad P_4 = z^4+192 tz.
\end{equation*}
Then the functions $ Q_n $  are regular, global in time solutions of NV at $ E = 0 $, localized as $ | z |^{ - 2 ( n + 1 ) } $ in space, such that $ \| Q_n( t ) \|_{ L^2( \mathbb{R}^2 ) } \underset{t \to \infty}{\to}\infty $. 

Apart from infinite time blow-up solutions, NV at $ E = 0 $ also possesses finite time blow-ups. The first explicit blow-up solution was constructed in \cite{TT}. This result was slightly generalized in \cite{KM}, more precisely it was noted that for any $a,c,d \in \mathbb{R} $ such that $a+c (x^3+y^3) +d(x^2+y^2)^2>0$ everywhere, the function
\begin{equation}\label{tai}
v(x,y,t)= -2 \Delta_{x,y} \log (a -24c t +c (x^3+y^3) +d(x^2+y^2)^2)
\end{equation}
where $\Delta_{x,y}$ is the 2d Laplacian with respect to $x$ and $y$, solves equation (\ref{NV}) with $ E = 0 $, decays like $r^{-3}$ at infinity ($r=\sqrt{x^2+y^2}$), and blows up at finite time if $ c \neq 0 $.

Numerically the formation of blow-up from initial data of the form of 
a ``KdV ring'' (initial data of the form $-0.5 \cosh 
(0.5(\sqrt{x^{2}+y^{2}}-20))$) for the NV equation at $ E = 0 $ was observed in \cite{Cetal} (\S 4.1).

In this paper we investigate, in particular, the possibility of a blow-up for NV equation at $ E \neq 0 $ for large enough initial data and for perturbations of the KdV line soliton.

\subsection{Travelling waves}
We say that $ v $ is a travelling wave if $ v( z, t ) = V( z - ct ) $ for some $ c \in \mathbb{C} $. We say that a travelling wave is a lump if it is weakly (algebraically) localized in space, as opposed to solitons which are exponentially localized in space.

The exact rational solutions for equation (\ref{NV}) at $ E > 0 $ 
were constructed by P.G. Grinevich and V.E. Zakharov, see \cite{G} 
(containing also a reference to a private communication from V.E. Zakharov). The solutions of this family are given by
\begin{equation}
\label{gz_potentials}
\begin{aligned}
& v( z, t ) = - 4 \partial_z \partial_{ \bar z } \ln \det A, \\
& w( z, t ) = 12 \partial_z^2 \ln \det A, \\
\end{aligned}
\end{equation}
where $ A = (A_{lm}) $ is $ 4 N \times 4 N $--matrix,
\begin{equation}
\label{gz_a_elements}
\begin{aligned}
& A_{ ll } = \frac{ i E^{ 1 / 2 } }{ 2 } \left( \bar z - \frac{ z }{ \lambda_l^2 } \right) - 3 i E^{ 3 / 2 } t \left( \lambda_l^2 - \frac{ 1 }{ \lambda_l^4 } \right) - \gamma_l, \\
& A_{ lm } = \frac{ 1 }{ \lambda_l - \lambda_m } \text{ for } l \neq m,
\end{aligned}
\end{equation}
and $ \lambda_1, \ldots, \lambda_{ 4 N } $, $ \gamma_1, \ldots, \gamma_{ 4N } $ are complex numbers such that
\begin{equation}
\label{gz_lambda_conditions}
\begin{aligned}
& \lambda_j \neq 0, \quad | \lambda_j | \neq 1, \quad j = 1, \ldots, 4N, \quad \lambda_l \neq \lambda_m \text{ for } l \neq m, \\
& \lambda_{ 2 j } = - \lambda_{ 2 j - 1 }, \quad \gamma_{ 2 j - 1 } - \gamma_{ 2 j } = \frac{ 1 }{ \lambda_{ 2 j - 1 } }, \quad j = 1, \ldots, 2N, \\
& \lambda_{ 4 j - 1 } = \frac{ 1 }{ \overline{\lambda}_{ 4 j - 3 } }, \quad \gamma_{ 4 j - 1 } = \bar \lambda_{ 4 j - 3 }^2 \bar \gamma_{ 4 j - 3 }, \quad j = 1, \ldots, N.
\end{aligned}
\end{equation}

The above formulas also define a solution of (\ref{NV}) if $ E < 0 $, but in the case $ E < 0 $ the above solutions have singularities in space (see \cite{FN}). As $ t \to \infty $ the Grinevich-Zakharov solution behaves as a sum of $ N $ travelling waves propagating with different velocities (see \cite{KN3}). 

If $ N = 1 $, then the above formulas give the following solution: 
\begin{gather}
v( z, t ) = - 8 \partial_z \partial_{ \bar z } \ln p_4, \text{ where } \\
p_4 = \kappa \bar \kappa + 2 \frac{ \lambda^2 \bar \lambda^2 + 1 }{ ( \lambda^2 \bar \lambda^2 - 1 )^2 }, \quad
\kappa = - \frac{ i \sqrt{ E } }{ 2 } \left( \bar z - \bar c t - 
\frac{ z - c t }{ \lambda^2 } + \gamma \right), \label{kappa}\\
c = 6E \left( \bar \lambda^2 + \frac{ 1 }{ \lambda^2 } + \frac{ \lambda^2 }{ \bar \lambda^2 } \right)
\end{gather}
and $ \lambda, \gamma \in \mathbb{C} $, $ \lambda \neq 0 $, $  | \lambda | \neq 1 $.
The set of admissible velocities $ c $ is given by Lemma 2.2 of \cite{KN2}. Note that if $ c \in \mathbb{R}_- $ and $ c $ belongs to the set of admissible velocities, i.e. $ c + 6E < 0 $, then the parameter $ \lambda $ can be calculated from $ c $:
\begin{equation}
\lambda^2 = - ( 1 + \tau ), \quad \tau = \frac{ \mu + \sqrt{ \mu^2 + 4 \mu }  }{ 2 }, \quad \mu = - \frac{ 6 E + c  }{ 6 E }.
\end{equation}
Thus one has always $\lambda^{2}<0$.

Note that the constant $\gamma$ in (\ref{kappa}) can be always 
absorbed by a translation  $z\mapsto z-z_{0}$. Thus in the case when $ c $ is real and $ c + 6E < 0 $,  putting $Z=z-ct$, we 
can write the lump in the form
\begin{equation}
    v(Z,t) = \left\{
    -\frac{E^{2}}{2}\left[\frac{1}{\lambda^{2}}\left(1+\frac{1}{\lambda^{4}}\right)(Z^{2}+\bar{Z}^{2})-\frac{4|Z|^{2}}{\lambda^{4}}\right]
    -4E\frac{(\lambda^{4}+1)^{2}}{(\lambda^{4}-1)^{2}\lambda^{4}}\right\}/p_{4}^{2}
    \label{lump}.
\end{equation}
The lump takes its global minimum of 
$-E(\lambda^{2}-1/\lambda^{2})^{2}$ for $Z=0$. 

It was shown in \cite{K2} that the rate of algebraic decay of Grinevich-Zakharov lumps is almost the strongest possible. More precisely, it was shown that the Novikov-Veselov equation (\ref{NV}) at nonzero energy does not possess travelling waves solutions decaying as $ O\left( | z |^{ - 3 - \varepsilon } \right) $, $ \varepsilon > 0 $, for $ | z | \to \infty $.

For $ E =0 $ the first regular lump for equation (\ref{NV_eq}) was 
constructed in \cite{Ch}. It reads as follows
\begin{align}\label{Chsol}
& v( z, t ) = U( \frac{ z }{ 2 }, t ), \quad w( z, t ) = - 3 V( \frac{ z }{ 2 }, t ), \text{ where } \\
& U( z, t ) = - \frac{ 240 z \bar z }{ ( 30 \bar z^2 z^2 + 1 )^2 }, \\
& V( z, t ) = \frac{ 3600 \bar z^4 z^2 - 120 \bar z^2 }{ ( 30 \bar z^2 z^2 + 1 )^2 }.
\end{align}
Note that this lump is stationary and is localized as $ |z|^{-6} $.

A generalization of this formula is given by (\ref{tai}). Note that formula (\ref{tai}) for $ c = 0 $ gives a stationary non-singular lump solution of NV at $ E = 0 $ provided that $ a,d>0 $. Another family of lump solutions for NV at $ E = 0 $ was constructed in \cite{KM2}:
\begin{equation}
\label{new_lumps}
v_{ a, b }( x, y, t ) = -2 \Delta_{ x,y } \log( 1 + ax +by +x^2 + y^2 ) =  \frac{ -2 ( 4 - a^2 -b^2 ) }{ ( 1 + ax +by +x^2 + y^2 )^2 }
\end{equation}
with $ a^2 + b^2 < 4 $ for smoothness. Note that these lumps are localized as $ |z|^{-4} $.

To our knowledge the question of stability of lump solutions has not 
yet been addressed for the NV equation (mainly because of their weak space localization). In this paper we present numerical results which might give an insight into this problem.

Note that the KdV soliton is a traveling wave solution to NV. For the 
closely related KP equation the question of stability of the KdV 
soliton was studied extensively, both numerically and theoretically 
(see, for example, references in \cite{KS1,KS2}). For NV at $ E = 0 $ it was shown numerically in \cite{Cr} that the KdV soliton is unstable with respect to perturbations periodic in the transverse direction. In the present article we study numerically the stability of KdV soliton with respect to localized perturbations for different levels of energy $ E $.

For nonlocalized travelling waves solutions or for travelling wave solutions having singularities see \cite{Cetal} and references therein.

\subsection{Large time behaviour}
It was shown in \cite{KN1} that the solutions of the Cauchy problem for the Novikov-Veselov equation at $ E > 0 $ corresponding to sufficiently regular, localized ``transparent'' initial data satisfying a small norm condition decay with time in $ L^{\infty} $ norm not slower than $ \frac{ 1 }{ t } $. (``Transparent'' potentials for the 2d Schr\"odinger equation (\ref{schrodinger}) are potentials $ v $ for which the associated scattering amplitude is identically zero (see \cite{KN1} for details). Note that, as shown in \cite{N2}, the localized travelling wave solutions for NV at $ E > 0 $ are transparent.)

In the case $ E < 0 $ it was shown in \cite{K1} that the solutions of the Cauchy problem corresponding to sufficiently regular, localized small norm initial data decay with time in $ L^{\infty} $ norm as $ \frac{ 1 }{ t^{ 3/4 } } $.

Note that the small norm assumption of the above papers prevents the formation of non-trivial traveling waves in the large time asymptotics. It would be interesting to study whether in a more general setting the ``soliton resolution conjecture'' is true for NV, i.e. whether for large times the solution decomposes into a sum of non-interacting traveling waves. 

\section{Numerical approaches and dynamic rescaling}
\label{dyn_resc}
In this section we will briefly review the used numerical approaches 
in this paper which are mainly identical to what has been 
successfully applied to numerical studies of KP solutions before. The 
idea is basically to consider the problems on $\mathbb{T}^{2}$ instead of 
$\mathbb{R}^{2}$, i.e., in a doubly periodic 
setting. The spatial dependence of the solutions will be treated by 
discrete Fourier transforms, the time dependence by fourth order 
exponential integrators. The numerical accuracy will be controlled by 
the conserved quantities of the NV equation which will numerically 
depend on time due to unavoidable numerical errors. 

Blow-up is expected to be self similar which allows an approach as 
for generalized KdV and generalized KP equations in terms of a 
dynamical rescaling. This will be presented here for the NV equation. 
Known exact blow-up solutions will be discussed within this 
framework.

\subsection{Numerical approaches}
The numerical task is to solve efficiently the NV equation 
(\ref{NV_eq}) which reads in Fourier space ($\xi=\xi_{1}+i\xi_{2}$)
\begin{equation}
    \partial_{t}\hat{v}=-i(\bar{\xi}^{3}+\xi^{3})\hat{v}\left(1-\frac{3E}{\xi\bar{\xi}}\right)
    +\bar{\xi}\widehat{vw} 
    +\xi\widehat{v\bar{w}},\quad 
    \hat{w}=-3\frac{\bar{\xi}}{\xi}\hat{v}.
    \label{NVfourier}
\end{equation}
The Fourier transforms in (\ref{NVfourier}) are approximated in 
standard way via discrete Fourier transforms, essentially truncated 
Fourier series. It is known that these \emph{spectral} methods are 
especially efficient when used to approximate analytic functions: the 
numerical error decreases in such cases exponentially with the number 
of Fourier modes, i.e., the number of terms in the truncated Fourier 
series. To take advantage of this so called spectral convergence of 
the method, we restrict our analysis to smooth periodic functions or to 
functions in the Schwartz space of rapidly decreasing smooth 
functions. The latter can be periodically continued on sufficiently 
large domains as functions analytic within the finite precision of 
the numerical approximation. 

Equation (\ref{NVfourier}) with discrete Fourier transforms instead 
of the standard Fourier transform is of the form $$
\partial_{t}\hat{v}=C\hat{v}+\mathcal{N}(\hat{v}),$$
where $C$ is a diagonal matrix, and where $\mathcal{N}$ denotes a nonlinear (and possibly 
nonlocal) function of $\hat{v}$. The eigenvalues of $C$ can have large numerical 
values of the modulus, whereas $\mathcal{N}$ is such that its $L^{\infty}$ norm is 
much smaller than these large eigenvalues. In other words the 
\emph{stiffness}\footnote{A system is called \emph{stiff} if it 
contains scales of vastly different orders which makes the use of 
standard explicit time integration schemes inefficient for stability reasons.} of this system is in the linear part which makes it 
suitable for time integration schemes as in \cite{KT} and references 
therein adapted to stiff systems. In \cite{KR11}, different stiff 
integrators were compared for KP equations. It was found that 
\emph{exponential integrators} perform best for this equation, and 
that the performances of such different integrators of fourth order 
are very similar. Thus we apply here the method of Cox and Matthews 
\cite{CM}.  The so called $\phi$ functions appearing in this approach 
are computed as in \cite{KT} and in \cite{Schm} with complex contour 
integrals. 

The numerical accuracy is controlled in two ways: first the Fourier 
coefficients are traced during the computations. It is well known that 
the Fourier coefficients for an analytic function decrease 
exponentially, and that the order of magnitude of the highest 
appearing wave numbers indicate the numerical error due to the 
truncation. This allows to determine the spatial resolution during 
the computation. It is always chosen in a way that the Fourier 
coefficients for the initial data decrease to machine precision (here 
$\sim 10^{-16}$), and the code is stopped once the Fourier coefficients 
near a blow-up no longer decrease below $10^{-3}$. The resolution in 
time is controlled as in \cite{etna,KR11} via the conserved quantities 
(\ref{H_energy}). Due to unavoidable numerical errors, the 
numerically computed such integrals will depend on time. We consider 
here
\begin{equation}
    \Delta=\left|1-\frac{H(t)}{H(0)}\right|
    \label{delta}.
\end{equation}
It was shown 
in \cite{KR11} that the relative conservation of such quantities 
overestimates the numerical accuracy in the $L^{\infty}$ norm by 1 to 
2 orders of magnitude. Near a blow-up, the code is stopped once the 
relative conservation of (\ref{H_energy}) is no longer better than $10^{-3}$. In 
this  case, the accuracy of the solution should be still of the order 
of plotting accuracy. 

\subsection{Dynamic rescaling}

In \cite{KP13} we have used a dynamic rescaling of the generalized KP equations 
to analyze blow-up in more detail with an adaptive approach. This 
method can be also applied to NV equations. The basic idea is to 
use the scaling invariance (\ref{invariance}) of the NV equation, 
but now with a time dependent scaling factor. 
As for generalized KP we consider the 
coordinate change with a factor $L(t)$
\begin{equation}
    \zeta = \frac{z-z_{m}}{L},\quad \xi = \frac{x-x_{m}}{L},\quad 
    \eta= \frac{y-y_{m}}{L},\quad
    \frac{d\tau}{dt}=\frac{1}{L^{3}},\quad V = L^{2}v, \quad W = 
    L^{2}w
    \label{gKP4},
\end{equation}
where $z_{m}(t)=x_{m}(t)+iy_{m}(t)$.
This leads for (\ref{NV_eq}) to
\begin{equation}
    \partial_{\tau}V=a\left(2V+\xi \partial_{\xi}V+\eta 
    \partial_{\eta}V\right)+2\Re(c\partial_{\zeta}V)+4\Re 
    \left(4\partial^{3}_{\zeta}V+\partial_{\zeta}(VW)-EL^{2}\partial_{\zeta}
    W\right)
    \label{gKP5},
\end{equation}
with 
\begin{equation}
    a=\partial_{\tau}(\ln L),\quad 
    c=\frac{\partial_{\tau}z_{m}}{L}.
    \label{a}
\end{equation}
Since $L$ will tend to zero at blow-up, the space and time scales are changed 
adaptively around the critical point which is reached here for $\tau\to\infty$. 
In contrast to generalized KP, there is no asymmetry in $x$ and $y$ with respect 
to the rescaling with $L$. Assuming that there is a self similar 
blow-up in certain NV solutions, the above rescaling suggests that  
asymptotically for large $\tau$ both $V$ and $W$ and $a$, $c$ become $\tau$ 
independent, which will be denoted with a superscript $\infty$. In 
this case equation (\ref{gKP5}) takes the form
\begin{equation}
    0=a^{\infty}\left(2V^{\infty}+\xi \partial_{\xi}V^{\infty}+\eta 
    \partial_{\eta}V^{\infty}\right)+2 \Re( c^{\infty}\partial_{\zeta}V^{\infty})+4\Re 
    \left(4\partial^{3}_{\zeta}V^{\infty}+\partial_{\zeta}(V^{\infty}W^{\infty})\right)
    \label{inf}.
\end{equation}
Solutions to (\ref{inf}) would give the blow-up profile of the self 
similar blow-up if solutions which are vanishing for $|\zeta| \to \infty $ exist. Note that there is no more dependence on $E$ in this 
equation which implies that such a blow-up profile would not depend 
on $E$.   Equation (\ref{inf}) is very different 
depending on whether $a^{\infty}$ is zero or not. The former case 
corresponds to an algebraic dependence of $L$ on $\tau$ which has 
been observed in generalized KdV systems in  the $L^{2}$ critical case 
\cite{MMR}. In this case, we have $L(\tau) = 
C_1\tau^{\gamma_1}$ with constants $C_{1}$, $\gamma_{1}$ with $\gamma_1<-1/3$ and thus 
$a^{\infty}=0$ as well as
\begin{equation}
   L(t) \propto (t^* - t)^{1/(3+1/\gamma_{1})}.
   \label{eq:Crit_Lt}
\end{equation}
This implies
\begin{equation}
    ||v||_{\infty}\propto ||v_{x}||_{2}\propto \frac{1}{(t^* - t)^{2/(3+1/\gamma_{1})}}
    \label{critnorm}.
\end{equation}
Equation (\ref{inf}) reduces 
to the equation for travelling wave 
solutions of NV equations for $E=0$. 

In the $L^{2}$ supercritical case, an exponential dependence of 
$L$ on $\tau$ is expected for generalized KdV equations, but so far not proven. 
For exponential decay we have $L(\tau) = C_2 e^{a^{\infty}\tau}$ with 
$C_{2}=const$ and $a^{\infty}<0$. Relation (\ref{gKP4}) implies 
in this case
\begin{equation}
   L(t) \propto (t^* - t)^{1/3},
   \label{eq:SupCrit_Lt}
\end{equation}
which leads to 
\begin{equation}
    ||v||_{\infty}\propto ||v_{x}||_{2}\propto \frac{1}{(t^* - t)^{2/3}}
    \label{supcritnorm}.
\end{equation}
Note that the NV equation is $L^{2}$ subcritical, but blow-up can occur 
here in contrast to generalized KdV or KP equations since the
conserved quantities of NV do not provide a control of the  $L^{2}$ 
norm for small $|E|$. It is one of the goals of the numerical studies in this 
paper to look for examples of blow-up in this case, and to identify 
the mechanism of the blow-up by tracing the $L^{\infty}$ norm of the 
solution and the $L^{2}$ norm of $\partial_{x}u$ in dependence of 
time. Comparison of these norms with (\ref{critnorm}) and 
(\ref{supcritnorm}) should give indications of the type of the 
observed blow-ups. 

As an example for the dynamical rescaling,
we consider the blow-up solution (\ref{tai}). Writing 
it as 
\begin{equation}
    v(x,y,t) = 
    -2\Delta_{x,y}\log\left(\frac{a}{24c}-t+\frac{1}{24}(x^{3}+y^{3})
    +\frac{d}{24c}(x^{2}+y^{2})^{2}\right)
    \label{tai2},
\end{equation}
one recognizes that $t^{*}=a/(24c)$ and that $L = (t^{*}-t)^{1/3}$. 
With (\ref{gKP4}) we get
\begin{equation}
    V =  -2\Delta_{\xi,\eta}\log\left(1+\frac{1}{24}(\xi^{3}+\eta^{3})
    +L\frac{d}{24c}(\xi^{2}+\eta^{2})^{2}\right)
    \label{tai3}.
\end{equation}
Thus the blow-up of this exact solution would correspond to the 
supercritical blow-up of generalized KdV (\ref{eq:SupCrit_Lt}). 

The dynamically rescaled equation (\ref{gKP5}) could of course be 
numerically integrated as the NV equation. But for a similar 
situation for the generalized KP equation, it was shown in \cite{KP14,KP13} 
that the terms $+\xi \partial_{\xi}V+\eta 
    \partial_{\eta}V$ are numerically problematic for a Fourier 
    approach. Thus we integrate as in \cite{KP13} the non rescaled 
    equation (\ref{NVfourier}) and trace the $L^{\infty}$ norm of the 
    solution and the $L^{2}$ norm of the gradient of the solution. 
    The dynamical scaling is then introduced a posteriori through 
    formulas (\ref{critnorm}) or (\ref{supcritnorm}).

\section{Perturbation of the KdV soliton}
\label{soliton_stab}
In this section we consider localized perturbations of the KdV 
soliton which reads in the present case for $t=0$
\begin{equation}
    v_{s}=a \mbox{sech}^{2}(bx),\quad a = \frac{6E+c}{4},\quad 
    b=\sqrt{\frac{-a}{2}},\quad 6E+c<0.
    \label{kdvsol}
\end{equation}
The NV equation is solved in a frame comoving with the soliton 
following from equation (\ref{NV_eq}) via the Galilei transformation 
$x\mapsto x-ct$. This leads to (we use the same symbols for the 
coordinates to simplify the notation)
\begin{equation}
    \partial_t v -c\partial_{x}v= 4 \Re ( 4 \partial_z^3 v + \partial_z( v w ) - E \partial_{ z } w ). \\
    \label{NVgal}
\end{equation}
We will consider a perturbation of the form 
\begin{equation}
    v(x,y,0) = v_{s}(x)-\alpha x\exp(-x^{2}-y^{2})
    \label{pertur}
\end{equation}
for the initial data, i.e., we study localized perturbations here. 
The value of $\alpha$ is in general chosen to be of the order of $10\%$ of 
the amplitude of the soliton $a$. We will denote $ N_t $ the number of time steps, and $ N_x $, $ N_y $ the number of Fourier modes corresponding to $ x $, $ y $ respectively.

\subsection{The KPI limit: $E\gg1$}

\begin{figure}
\centering
    \includegraphics[width=0.45\textwidth]{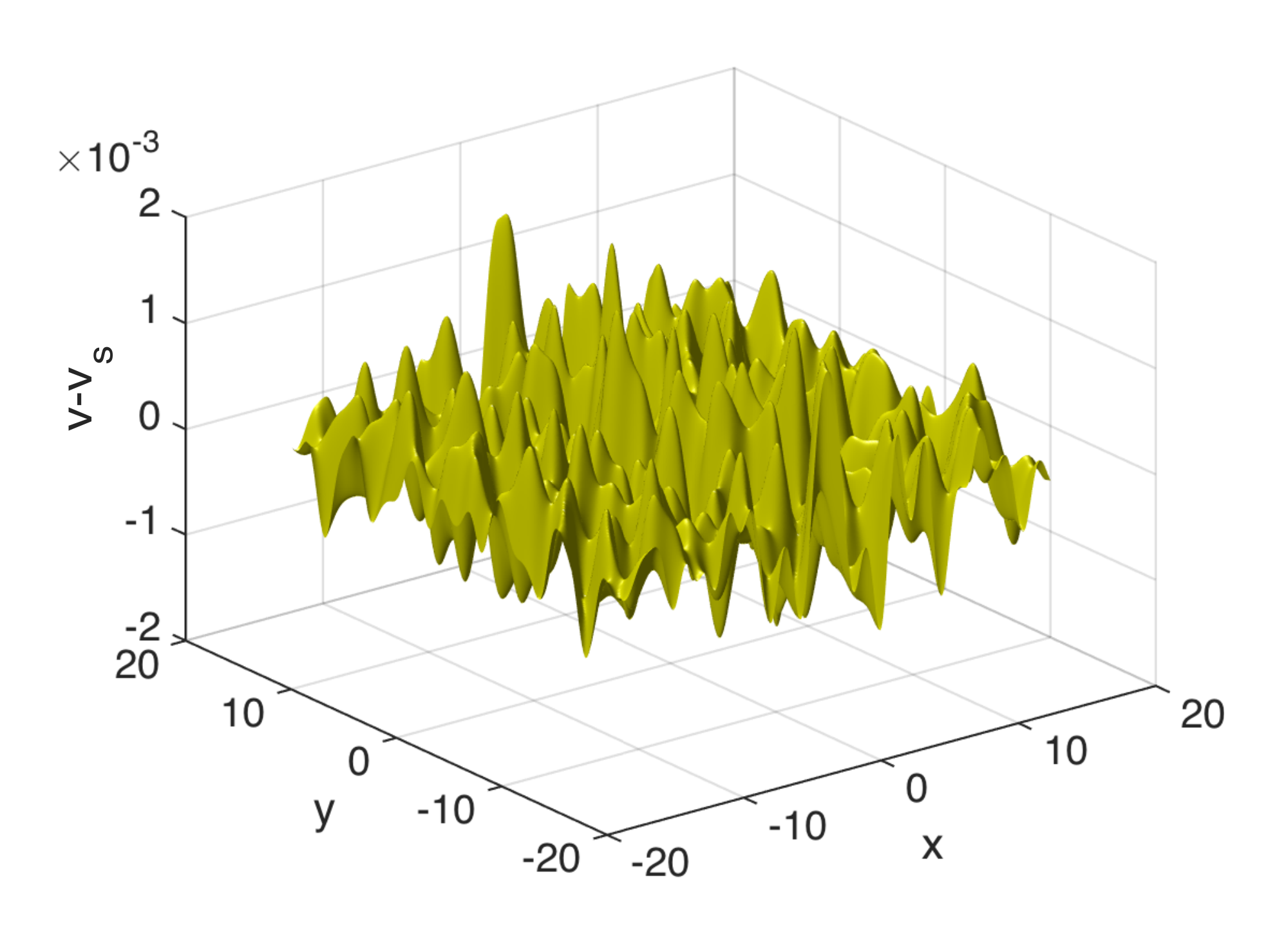}
    \includegraphics[width=0.45\textwidth]{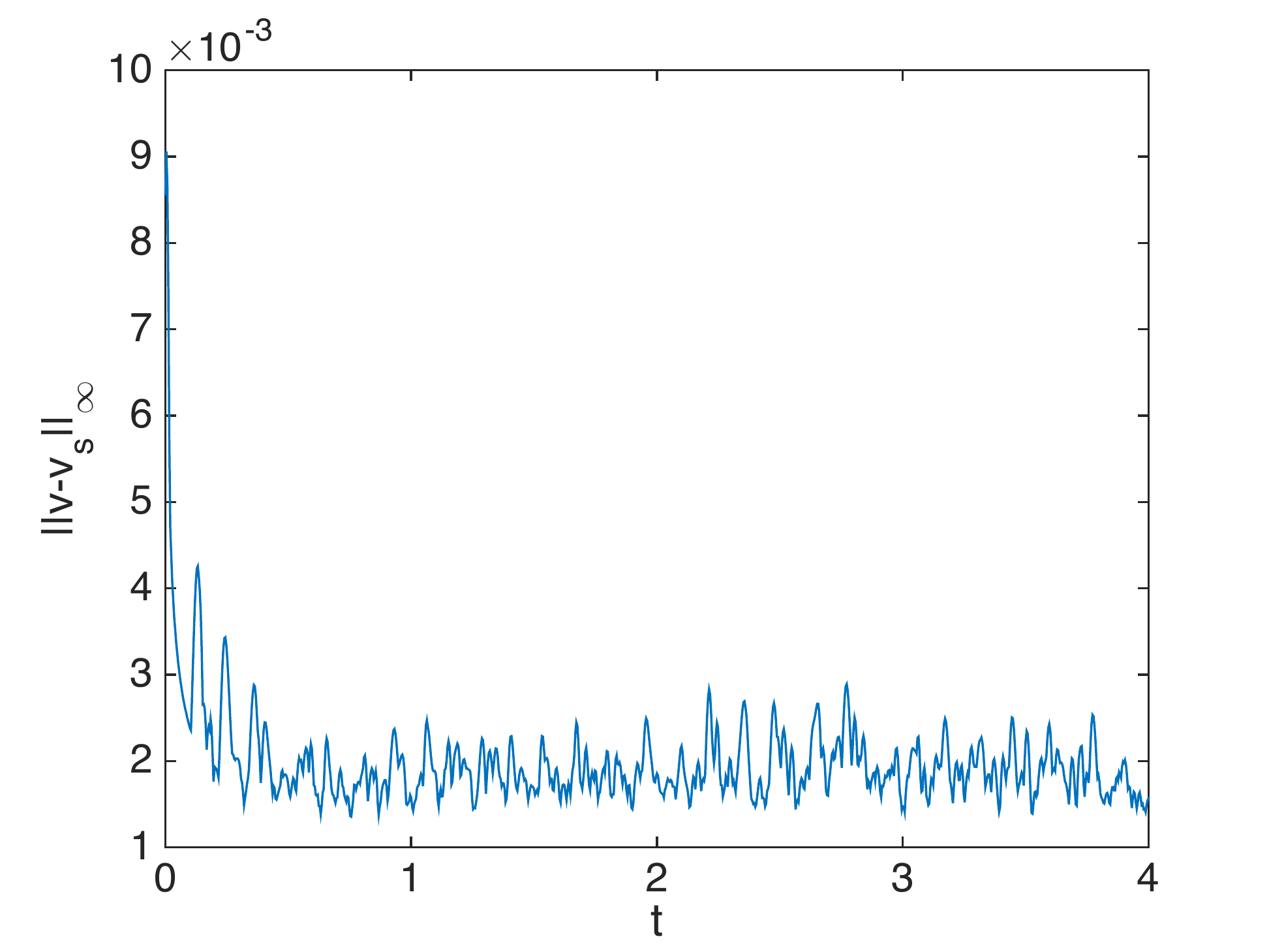}
 \caption{Difference between the solution to the NV equation (\ref{NVgal}) and $ v_s $
 with $E=10$ for the initial data (\ref{pertur}) with $c=-61$ and 
 $\alpha=0.02$ for $t=4$ on the left and the $L^{\infty}$ norm of the 
 difference $v-v_{s}$ in dependence of $t$ on the right. }
 \label{VNsolE10}
\end{figure}

We first consider the case of large positive $E$ ($E=10$) since it is known 
that the NV equation approaches the KP I equation in the limit 
$E\to+\infty$. Thus in 
this case we expect a qualitatively similar behavior as for the KP I perturbations studied in \cite{Zak,RT}. It 
is known that there is a value $a^{*}$ such that for $a<a^{*}$ the 
KdV soliton is non-linearly stable in KP I. Choosing $c=-61$, $\alpha=0.02$, 
$N_{t}=1000$ for $0<t<4$ and $N_{x}=2^{9}$, $N_{y}=2^{8}$, we get the 
situation shown in Fig.~\ref{VNsolE10}. The solution is fully 
resolved in Fourier space in the sense that the Fourier coefficients 
decrease to the order of $10^{-10}$, the numerically computed 
relative energy $ H $ (see (\ref{H_energy})) is conserved to the order of $\Delta\sim 10^{-10}$ throughout the 
computation. Since we study the problem in a periodic setting, the 
initial perturbation cannot be radiated away to infinity, which will 
be done mainly along rays with an angle of $2\pi/3$ between each 
other, but reenter on opposing sides of the computational domain. 
Thus the initial perturbation leads to an agitated background in the 
form of some almost random noise. In the same figure we show on the 
right the $L^{\infty}$ norm of the difference between numerical 
solution and the soliton in dependence of time. It can be seen that 
this norm quickly decreases to the level of this background. Thus 
there is no indication of an instability of the soliton in this case. 

\begin{figure}
\centering
    \includegraphics[width=\textwidth]{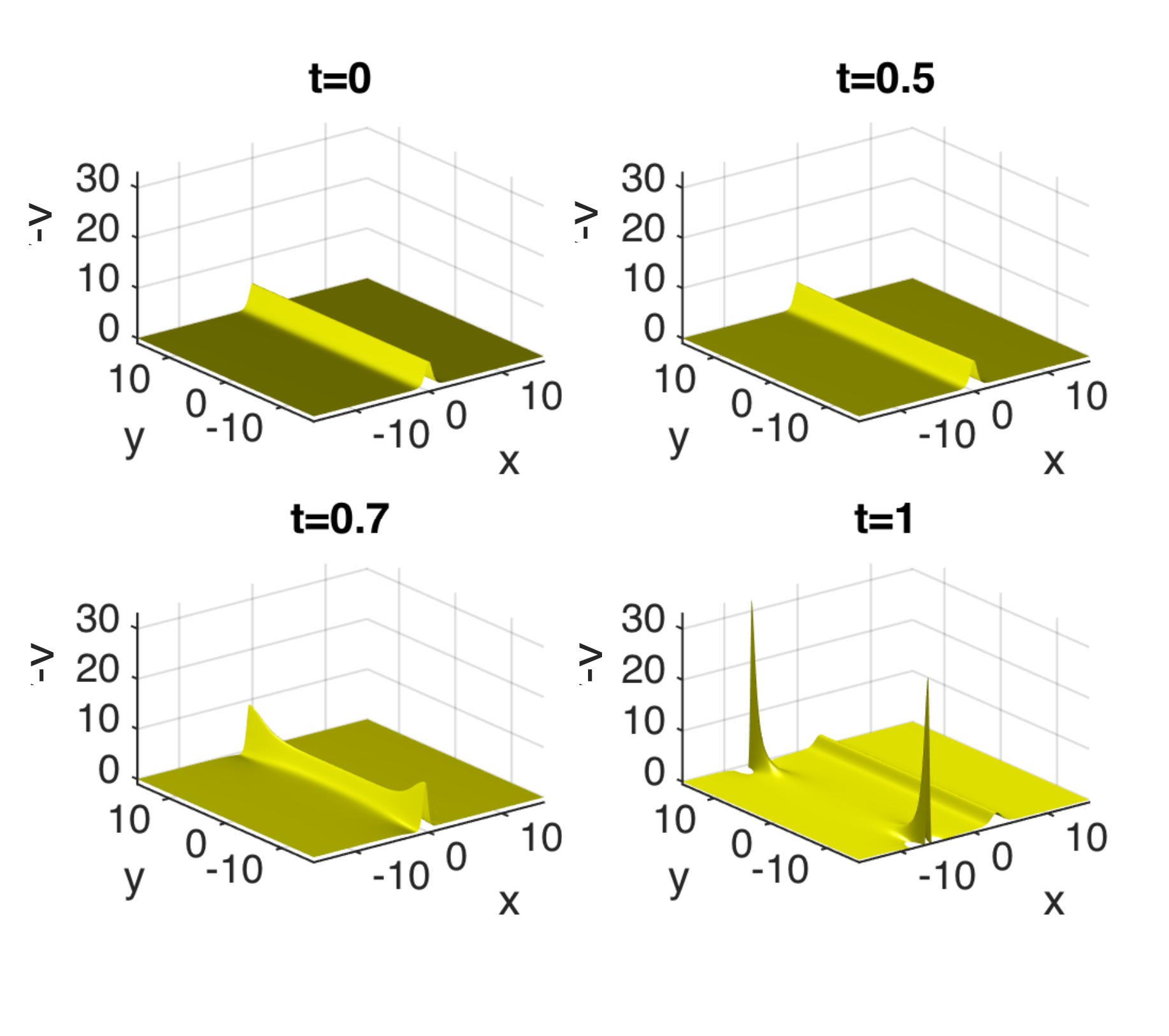}
 \caption{Solution to the NV equation (\ref{NVgal}) 
 with $E=10$ for the initial data (\ref{pertur}) with $c=-80$ and 
 $\alpha=0.02$ for different times.  }
 \label{VNsolE10cm80}
\end{figure}

The situation changes visibly if we consider the same situation as in 
Fig.~\ref{VNsolE10}, just this time with $c=-80$, thus a larger 
speed. The resulting solution can be seen at different times in 
Fig.~\ref{VNsolE10cm80}. The weak initial perturbation leads in this 
case to a growing of the solution at the boundaries in $y$ of the (periodic) 
computational domain. The solution eventually appears to develop into a lump. 

\begin{figure}
\centering
    \includegraphics[width=0.45\textwidth]{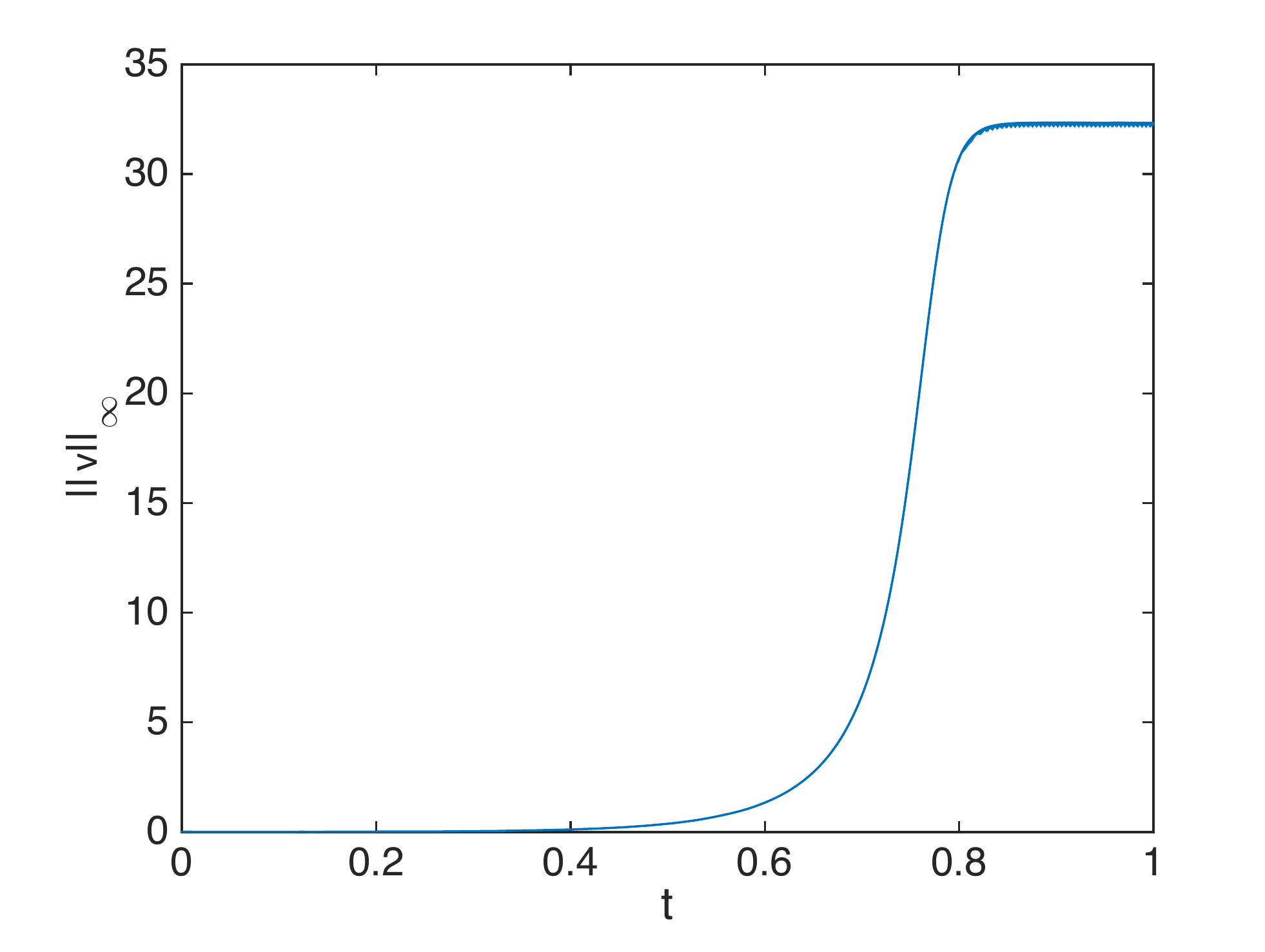}
    \includegraphics[width=0.45\textwidth]{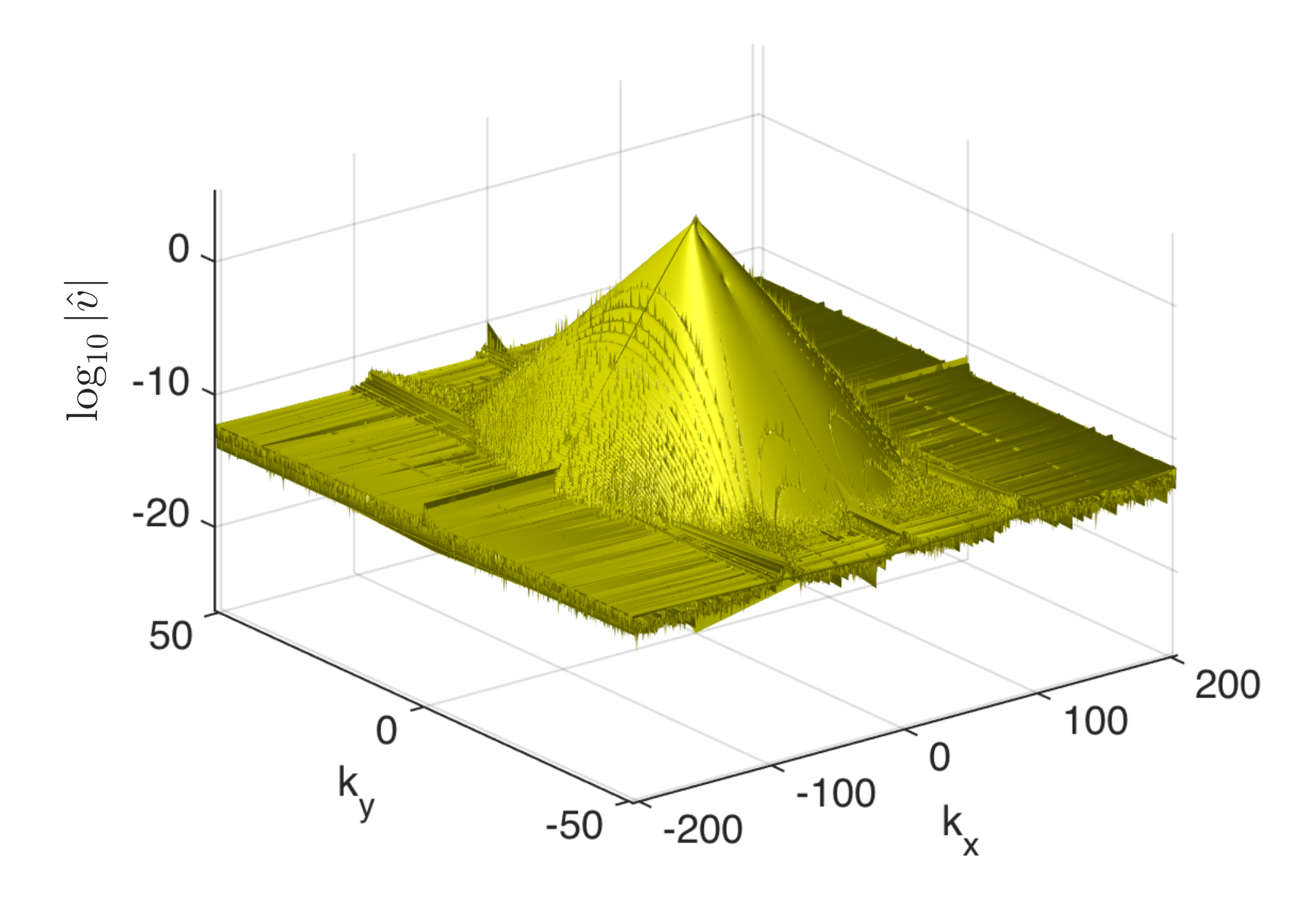}
 \caption{$L^{\infty}$ norm of the solution to the NV equation (\ref{NVgal}) 
 with $E=10$ for the initial data (\ref{pertur}) with $c=-80$ and 
 $\alpha=0.02$ in dependence on time on the left and modulus of the 
 Fourier coefficients of the solution for $t=1$ on the right. }
 \label{VNsolE10cm80fourier}
\end{figure}

The appearance of a lump is also confirmed by the $L^{\infty}$ 
norm of the solution which can be seen in 
Fig.~\ref{VNsolE10cm80fourier}. It reaches after some time a plateau 
which is generally the indication of the appearance of a lump. 
This implies that the KdV soliton is unstable as a solution to the NV 
equation, and also that the lump solution is stable.  Note that the 
solution is well resolved in Fourier space which can be seen on the 
right of Fig.~\ref{VNsolE10cm80fourier}. We used $N_{x}=2^{11}$ and 
$N_{y}=2^{9}$ Fourier modes and $N_{t}=10^{4}$ time steps. 
The numerically computed 
energy $ H $ is conserved relatively to the order of $10^{-4}$. 

\begin{figure}
\centering
    \includegraphics[width=0.7\textwidth]{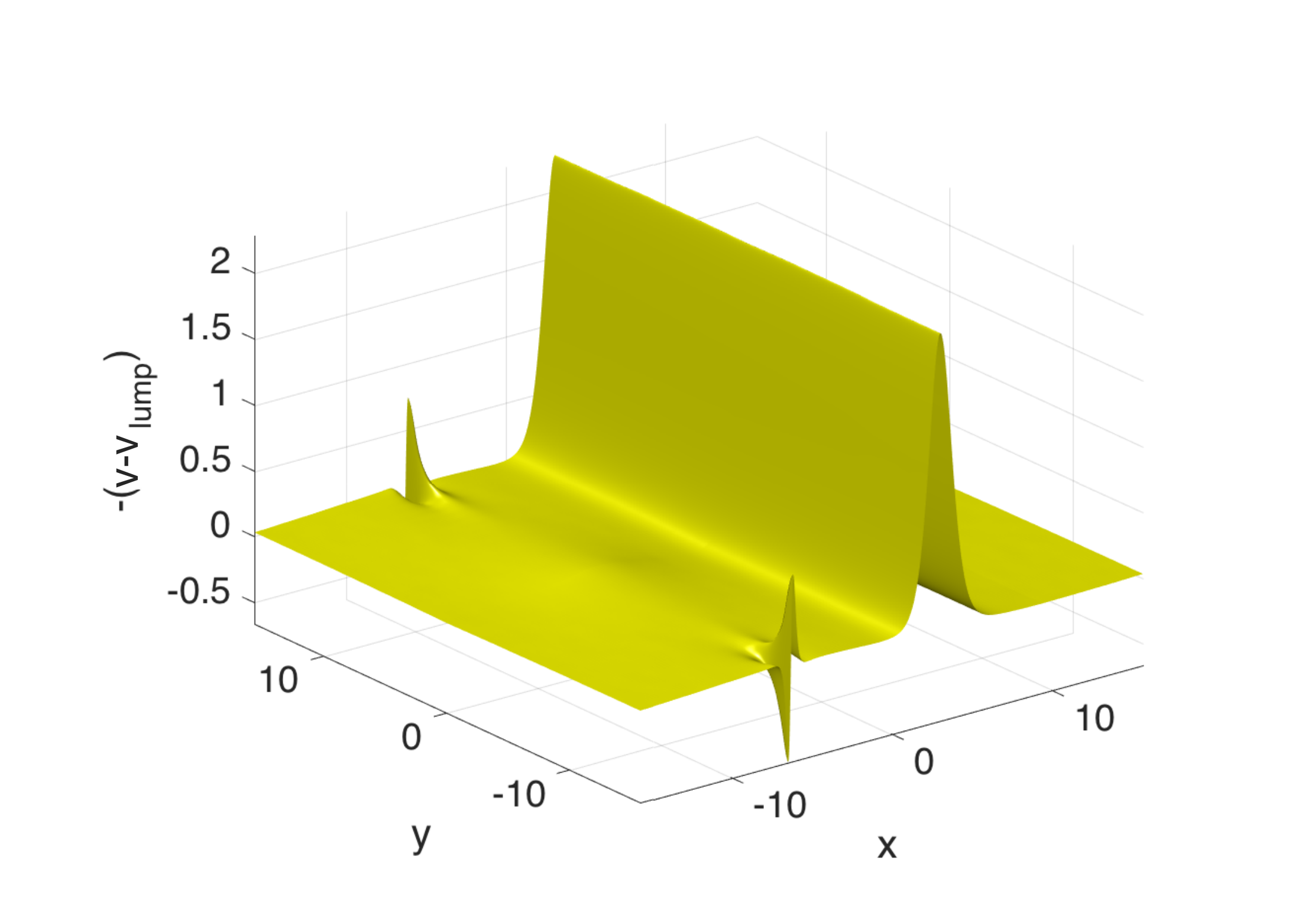}
 \caption{Difference between the solution to the NV equation (\ref{NVgal}) 
 with $E=10$ for the initial data (\ref{pertur}) with $c=-80$ and 
 $\alpha=0.02$ and a fitted lump (\ref{lump}) for $t=1$.  }
 \label{VNsolE10cm80lump}
\end{figure}

It is possible to fit the apparent lump to the 
formula (\ref{lump}). This is done by determining the parameter 
$\lambda$ in (\ref{lump}) in a way that the lump has the same minimal 
value as the solution at the peak. In Fig.~\ref{VNsolE10cm80lump} we 
show the difference between this lump centered at the location of the 
minimum of the solution to the NV equation in Fig.~\ref{VNsolE10cm80} 
in the last frame. It can be seen that the lump fits rather well, but 
that the solution is not yet sufficiently far from the remainder of 
the KdV soliton to be exactly given by the lump solution. The 
asymptotic final state appears to be lumps plus sufficiently small 
KdV solitons, which are stable. 

\subsection{The KPII limit: $E\ll -1$}

\begin{figure}
\centering
    \includegraphics[width=0.45\textwidth]{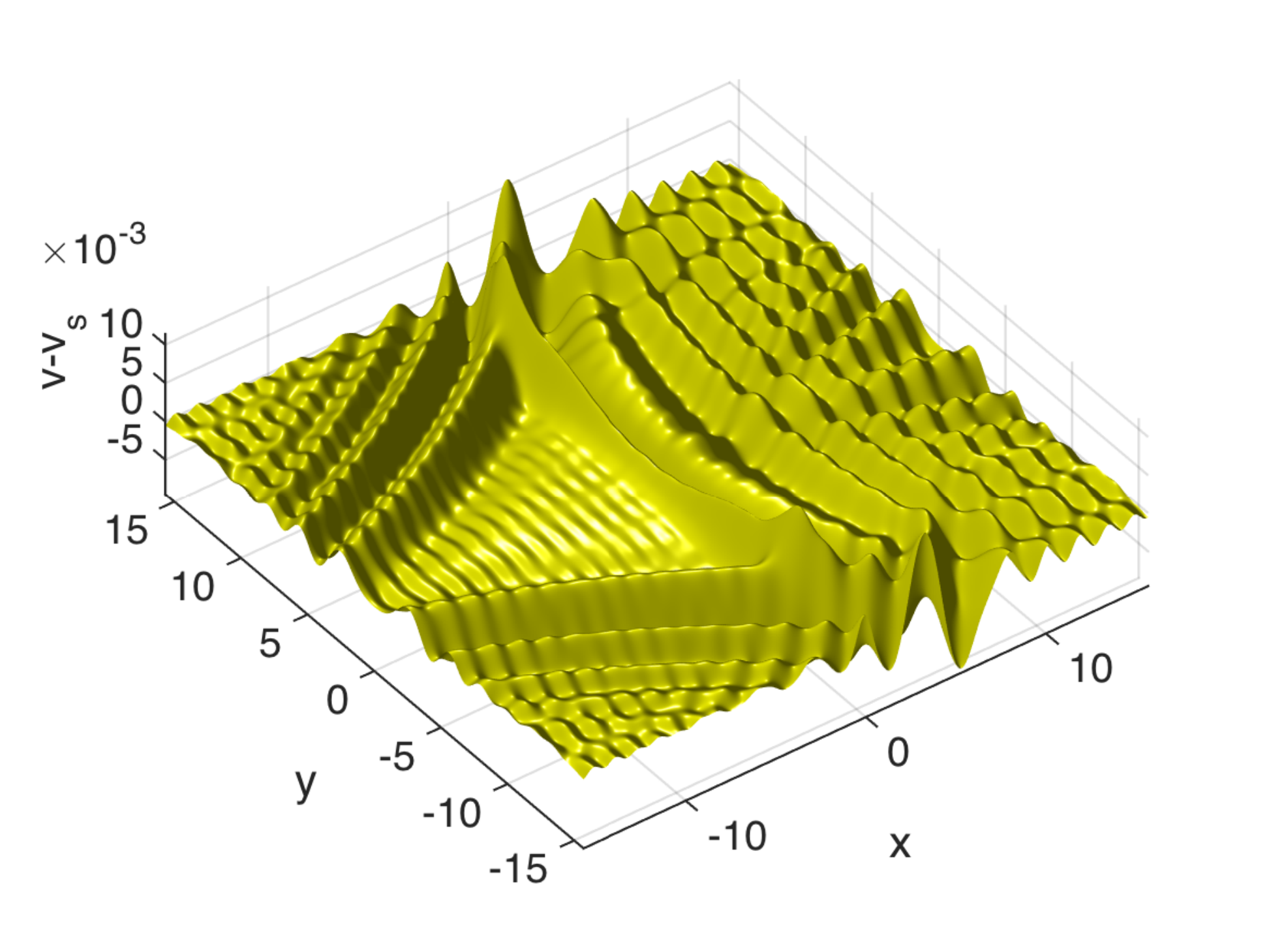}
    \includegraphics[width=0.45\textwidth]{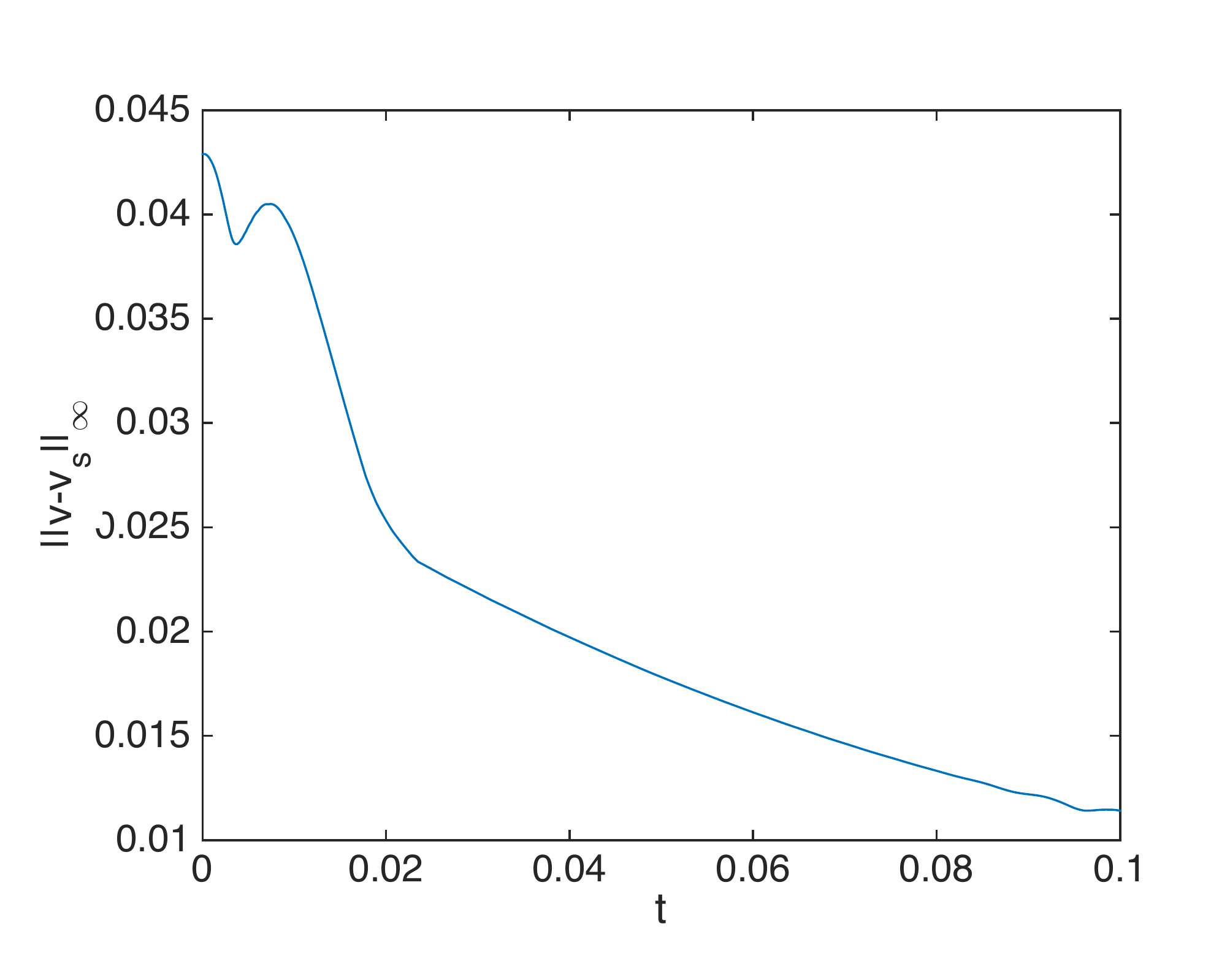}
 \caption{Difference between the solution to the NV equation (\ref{NVgal}) 
 with $E=-10$ for the initial data (\ref{pertur}) with $c=40$ and 
 $\alpha=0.1$ and the KdV soliton $v_s$ for $t=0.1$ on the left; dependence of the $L^{\infty}$ norm of $v-v_s$ on time on the right.}
 \label{VNsolEm10c40}
\end{figure}

The situation is different for large negative values of $E$, 
where the behaviour should be close to the KP II setting. Recall that 
the KdV soliton is stable for KP II. We consider $E=-10$, $c=40$ and 
$\alpha=0.1$, this time with $N_{x}=N_{y}=2^{10}$ and $N_{t}=10^{4}$ 
for $t\leq0.1$. 
In Fig.~\ref{VNsolEm10c40} we show the difference between the NV 
solution and the KdV soliton at $t=0.1$. It can be seen that the 
perturbation is radiated to infinity, again in the typical triangular 
pattern. Due to the imposed periodicity, the radiation reenters at 
the boundaries of the computational domain on the opposing side. The 
$L^{\infty}$ norm of the difference between NV solution and KdV 
solution in the same figure on the right also indicates that the 
soliton is stable and that the perturbation is simply radiated to 
infinity. 

\subsection{Intermediate values of $E$}

\begin{figure}
\centering
    \includegraphics[width=0.45\textwidth]{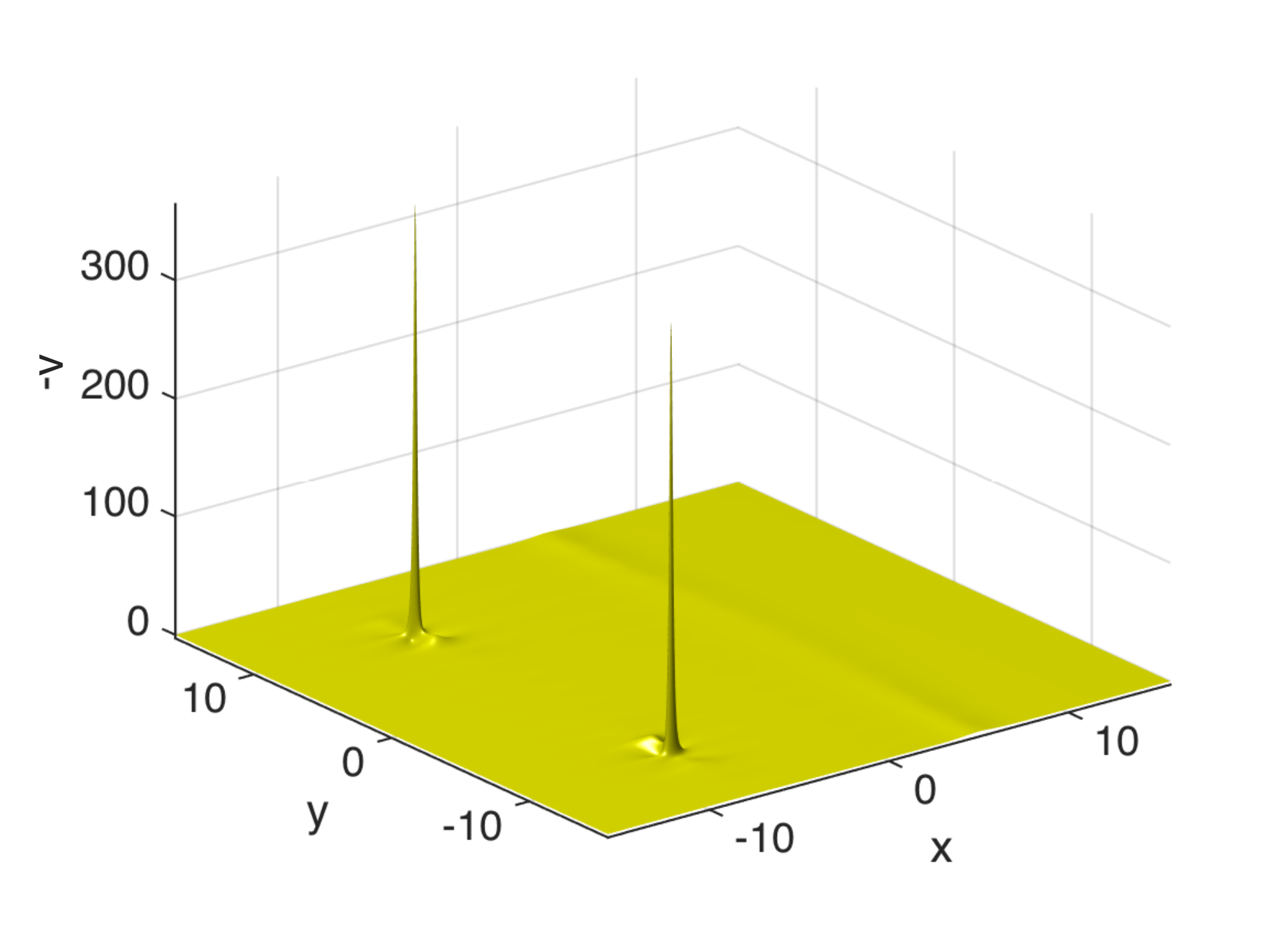}
    \includegraphics[width=0.45\textwidth]{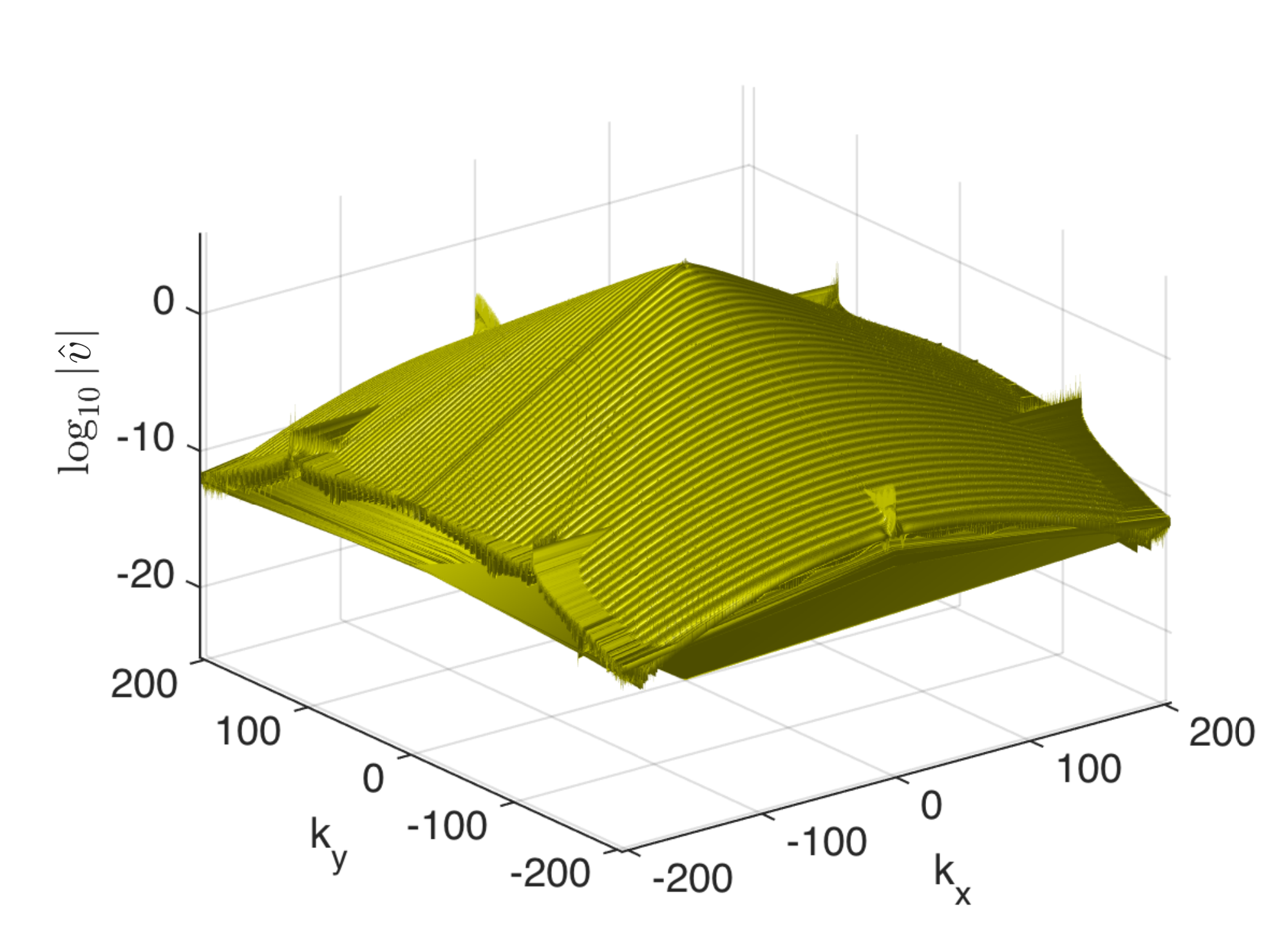}
 \caption{Solution to the NV equation (\ref{NVgal}) 
 with $E=1$ for the initial data (\ref{pertur}) with $c=-20$ and 
 $\alpha=0.1$ for $t=1.4925$ on the left and the corresponding Fourier 
 coefficients on the right. }
 \label{VNsolE1cm20}
\end{figure}

For a smaller positive value of $E$ ($E=1$), we find again that perturbations 
of a slow KdV soliton (small $|c|$) are stable. For larger values of 
the speed of the KdV soliton, once more an instability is observed. But 
this time, the appearing lump-like structures seem themselves to be 
unstable and to finally blow up. We again consider initial data of 
the form (\ref{pertur}), this time with $c=-20$ and $\alpha=0.1$. We 
use $N_{x}=N_{y}=2^{11}$ Fourier modes and $N_{t}=100000$ time steps. 
For 
$t> 1.49$, the  relative computed energy $H$ is no longer 
conserved to better than  $10^{-3}$ and the results are ignored. The 
solution at $t=1.4925$ can be seen in Fig.~\ref{VNsolE1cm20}. As can 
be seen in the same figure on the right, there is not yet a lack of resolution 
in the Fourier domain at this time. Thus the solution appears here 
to be underresolved in time. Note that generalized KP blow-ups generally are 
indicated by a lack of resolution in Fourier space. 

\begin{figure}
\centering
    \includegraphics[width=0.45\textwidth]{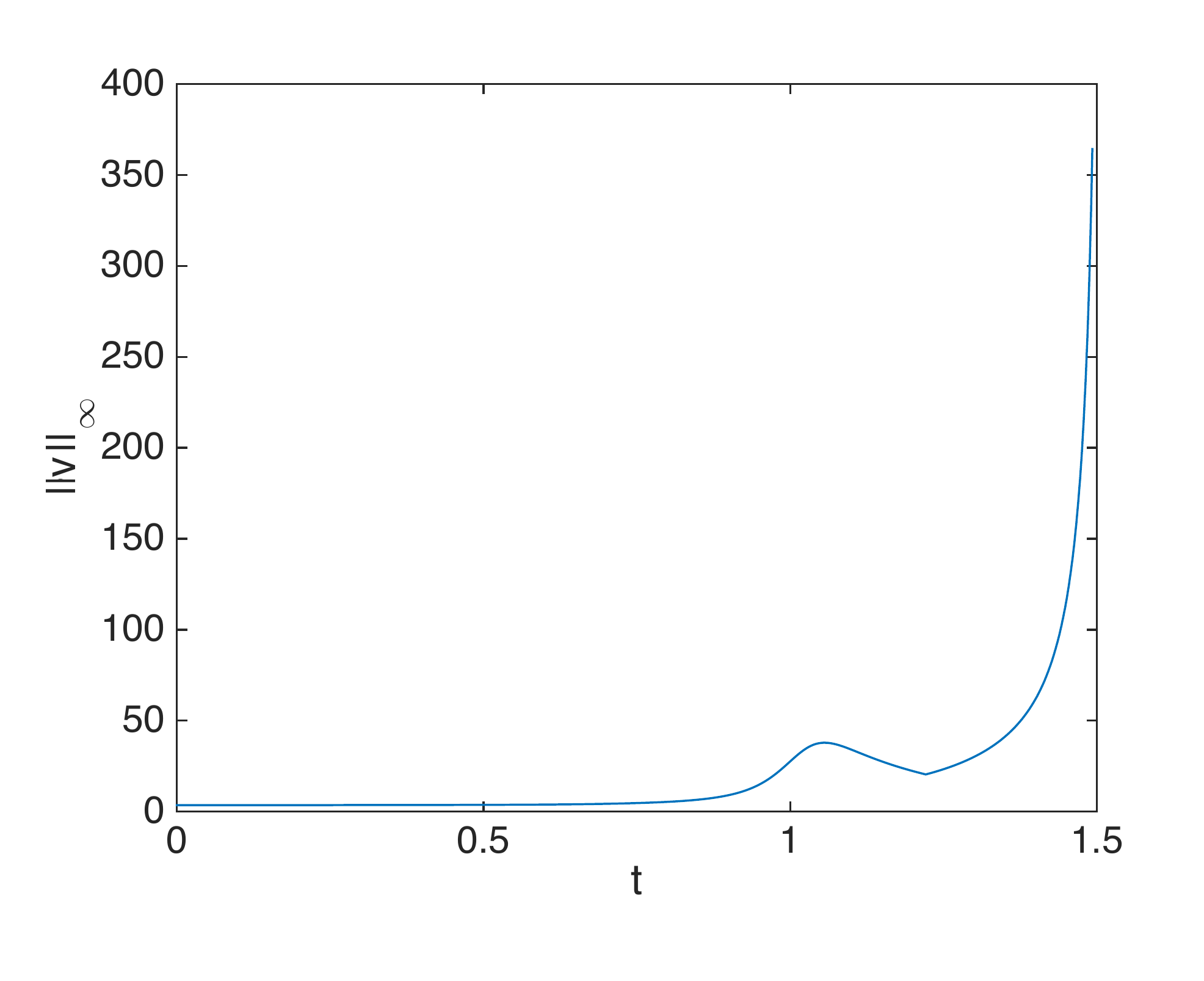}
    \includegraphics[width=0.5\textwidth]{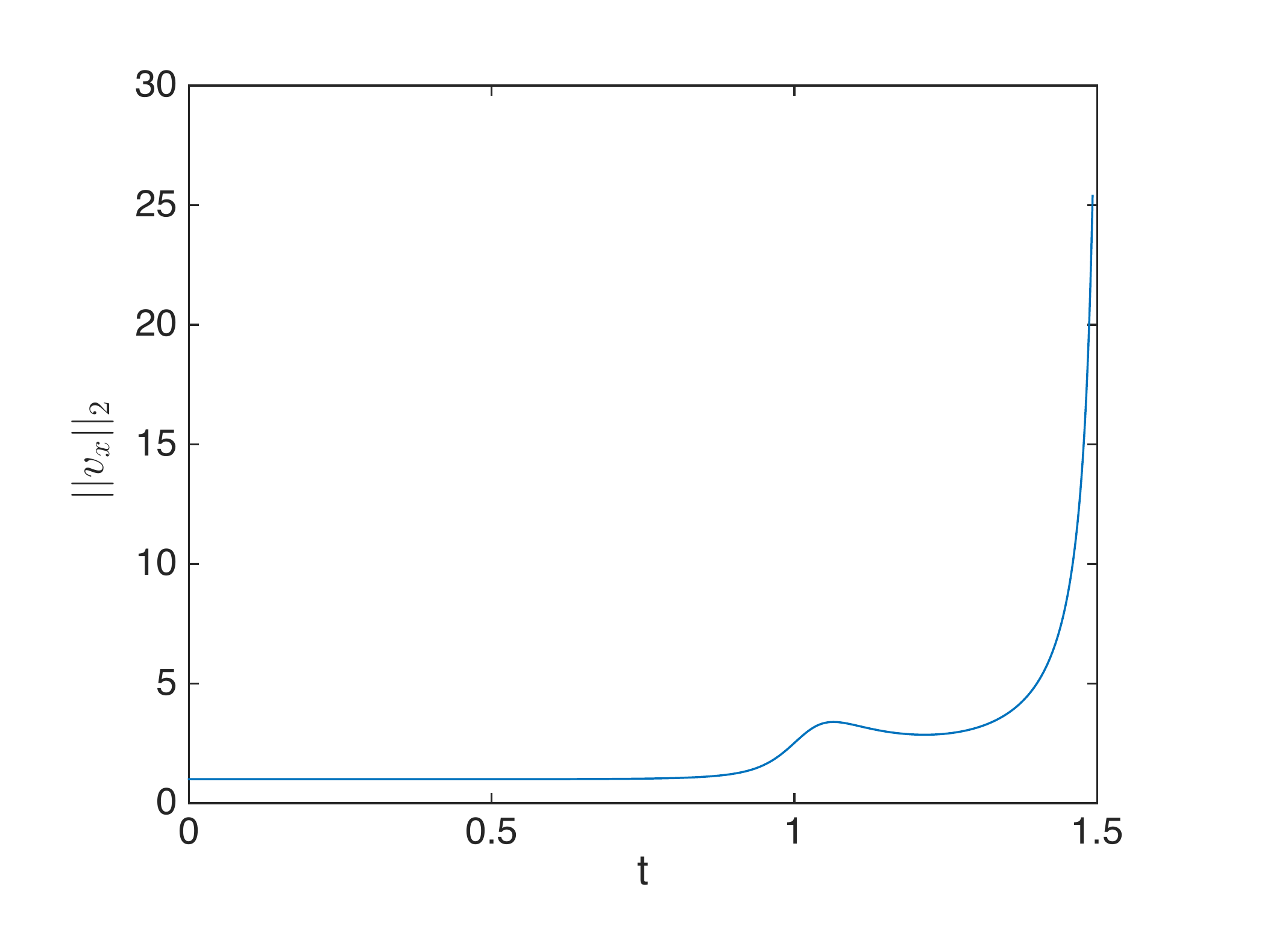}
 \caption{$L^{\infty}$ norm of the solution to the NV equation (\ref{NVgal}) 
 with $E=1$ for the initial data (\ref{pertur}) with $c=-20$ and 
 $\alpha=0.1$ in dependence on time on the left and the $L^{2}$ norm 
 of $v_{x}$  on the right. }
 \label{VNsolE1cm20norm}
\end{figure}

The $L^{\infty}$ norm of $v$ and the $L^{2}$ norm of $v_{x}$ both 
indicate a blow-up as can be seen in Fig.~\ref{VNsolE1cm20norm}. 
There appears to be a lump formation for $t\sim 1$ which then itself 
is unstable after a short time.  
        
\begin{figure}       
\centering
    \includegraphics[width=0.45\textwidth]{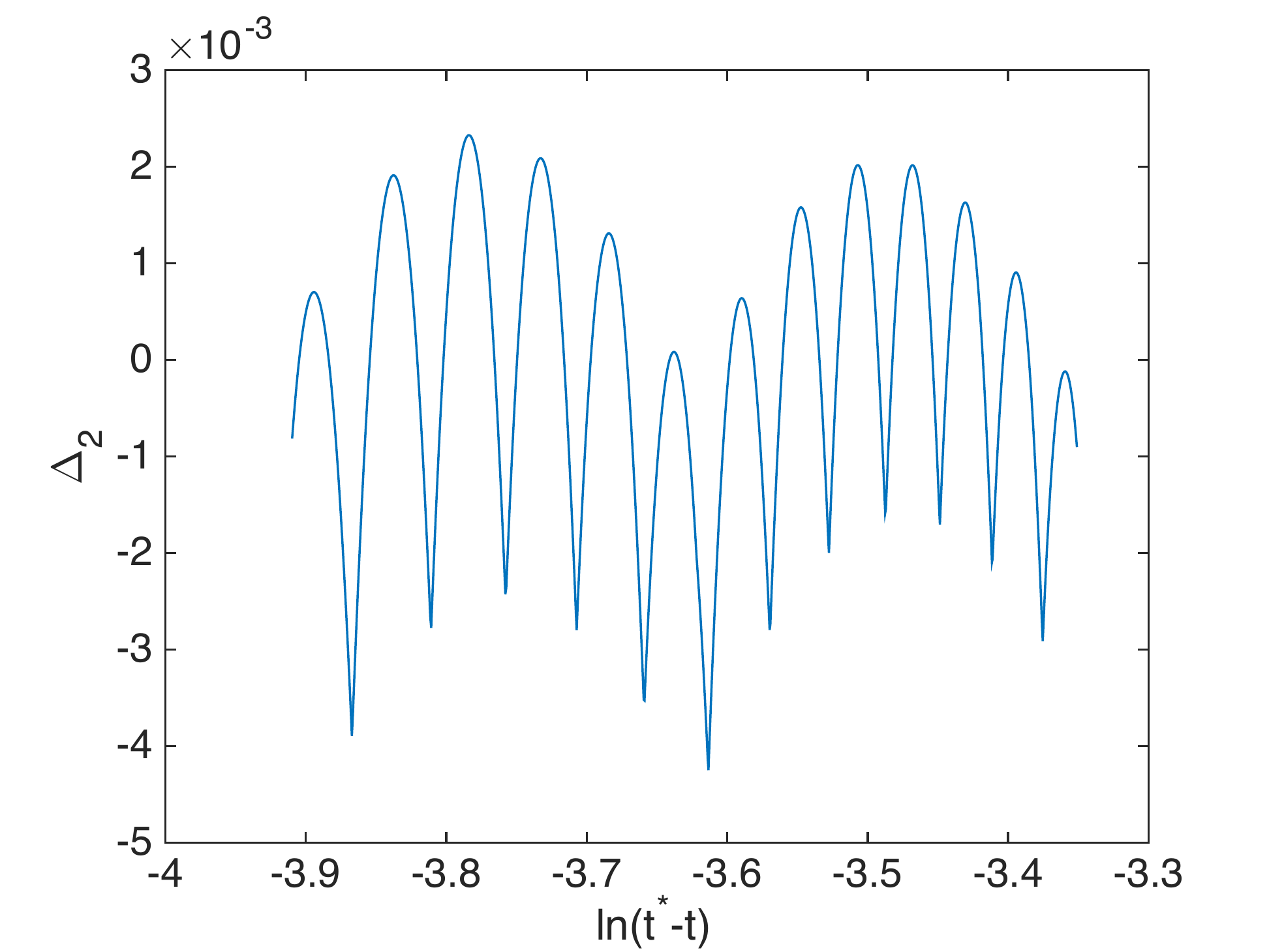}
    \includegraphics[width=0.45\textwidth]{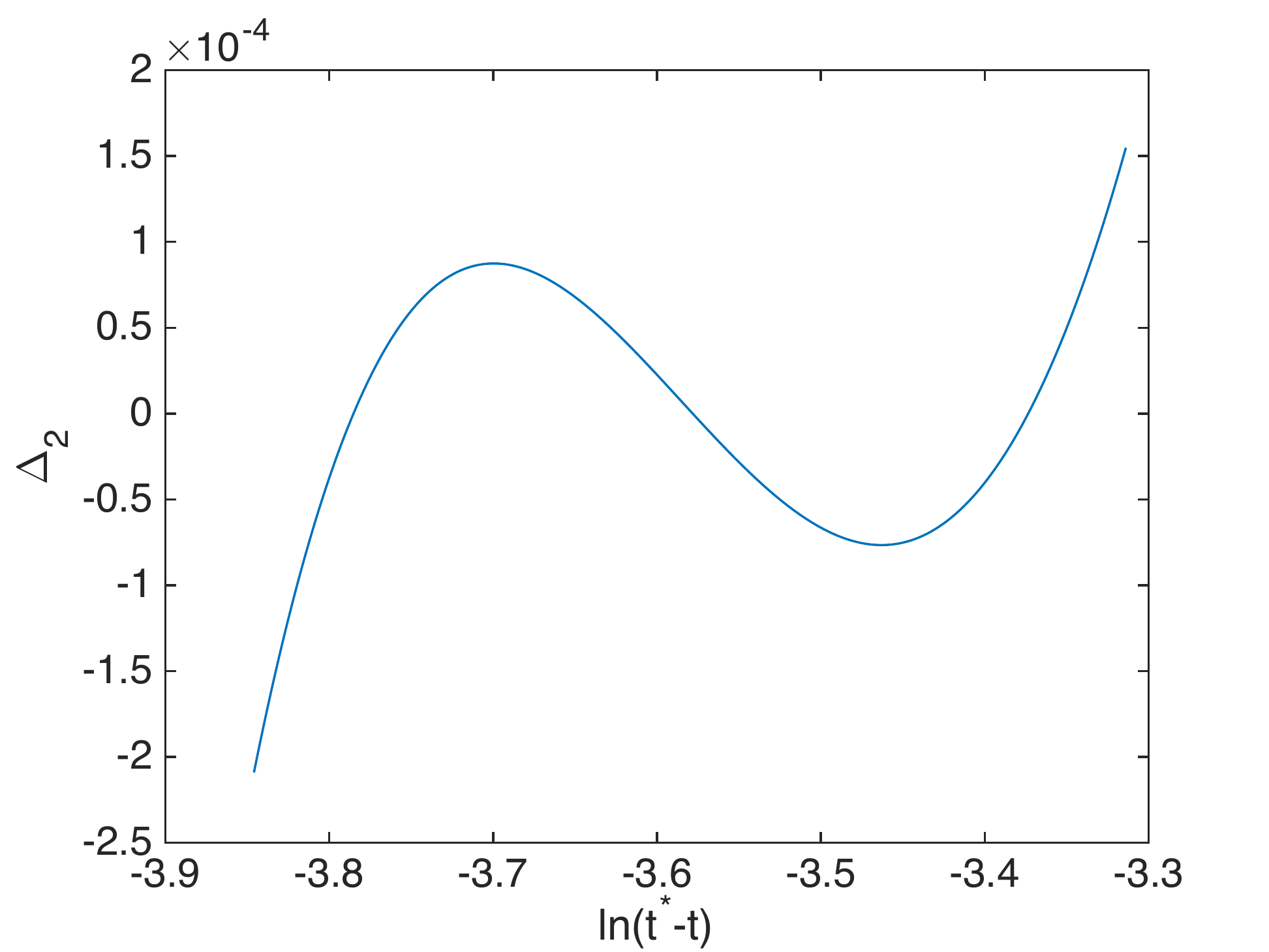}
 \caption{Fit of the norms in Fig.~\ref{VNsolE1cm20norm} to 
 $\gamma\ln (t^{*}-t)+\delta$; the quantity $\Delta_{2}:=|\ln 
 C-\gamma\ln (t^{*}-t)-\delta|$ for $C$ the $L^{\infty}$ norm of $v$ on the 
 left and for $C$ the $L^{2}$ norm of $v_{x}$ on the right. }
 \label{VNsolE1cm20fit}
\end{figure}

To understand the mechanism of a potential blow-up, we perform a fit 
of  $\ln C$ to $\gamma \ln (t^{*}-t)+\delta$, where $C$ is either the 
norm $||v||_{\infty}$ or the norm $||v_{x}||_{2}$, and where $ t^* $ is the blow-up time and $\gamma$ 
and $\delta$ are constants. This fit is performed for the last 1000 
time steps with the optimization algorithm \cite{fminsearch} which is 
distributed in Matlab as the command \emph{fminsearch}. For the 
example in Fig.~\ref{VNsolE1cm20} we get for the norm 
$||v||_{\infty}$ the values $t^{*}=1.5125$, $\gamma=-0.99$ and         
$\delta=2.04$, and for the norm $||v_{x}||_{2}$        
the values $t^{*}=1.5139$, $\gamma=-1.004$ and $\delta=-0.6261$. The 
quality of the fittings can be seen in Fig.~\ref{VNsolE1cm20fit}, the 
fitting errors are of the order of $10^{-3}$. Note that the results 
do not change much if the fitting is done for the last 500 time 
steps. The compatibility between the found blow-up times shows the 
consistency of the fitting. The results indicate that the factor $L$ 
in (\ref{gKP4})
should be proportional to $\sqrt{t^{*}-t}$, i.e. the mechanism
(\ref{eq:Crit_Lt}) with $\gamma_{1}=-1$. This would 
imply that the constant $a^{\infty}$ in (\ref{inf}) vanishes, and 
that the blow-up profile is the lump for $E=0$. 

\begin{figure}
\centering
    \includegraphics[width=0.45\textwidth]{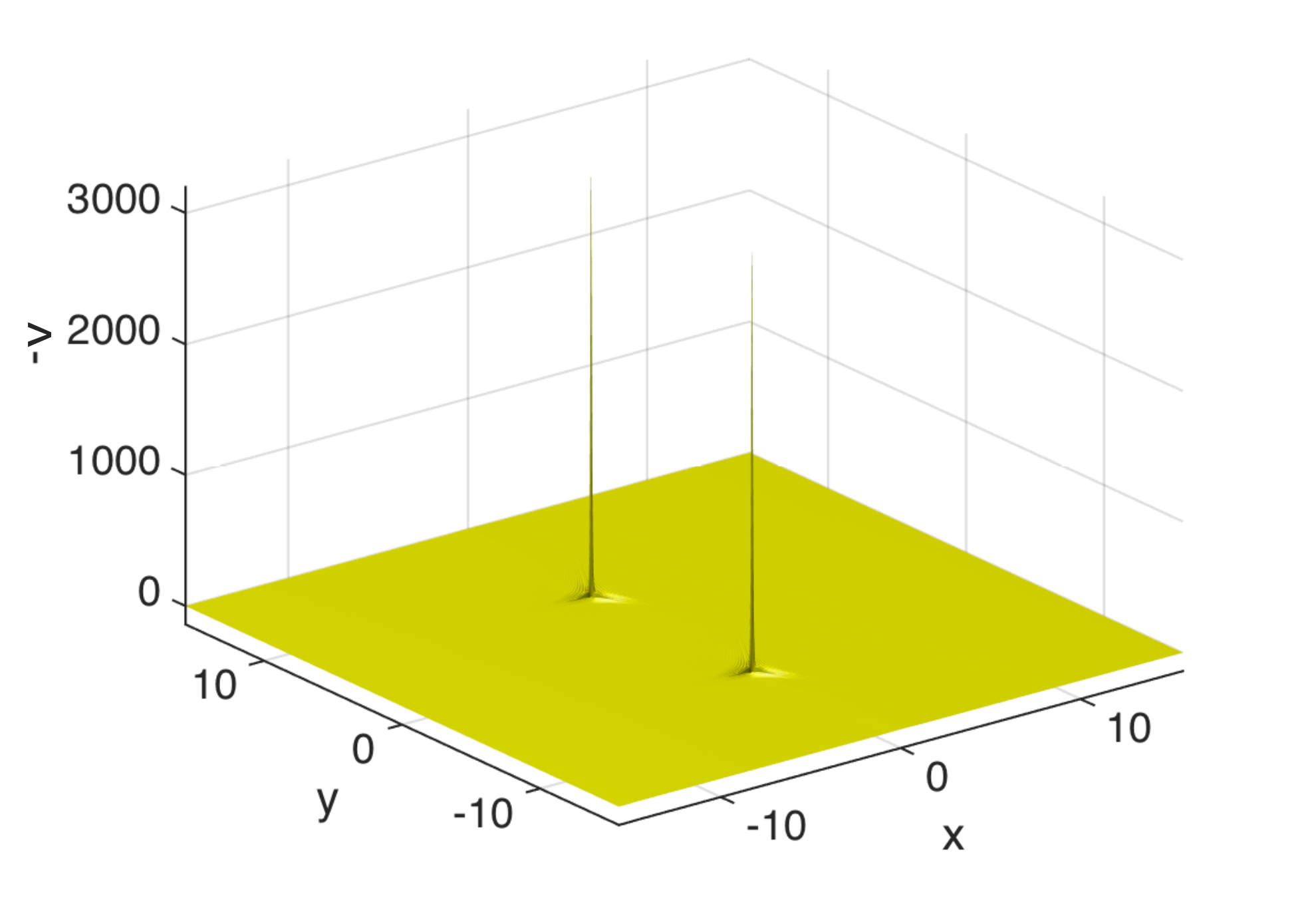}
    \includegraphics[width=0.45\textwidth]{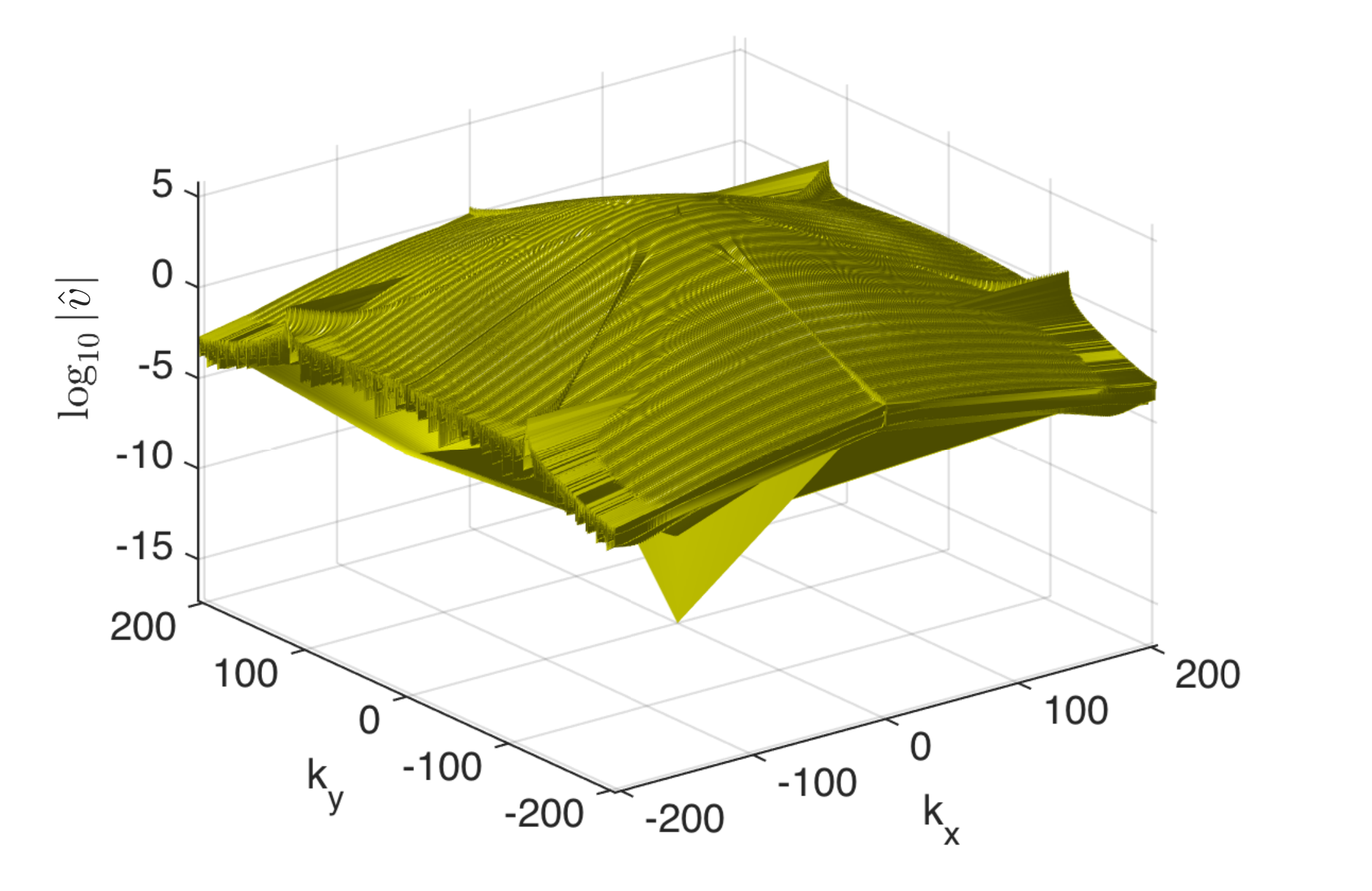}
 \caption{Solution to the NV equation (\ref{NVgal}) 
 with $E=0$ for the initial data (\ref{pertur}) with $c=-10$ and 
 $\alpha=0.1$ for $t=1.4775$ on the left and the corresponding Fourier 
 coefficients on the right. }
 \label{VNsolE0cm10}
\end{figure}

A similar picture is found for  $E=0$, perturbations 
of a slow KdV soliton (small $|c|$) are stable, whereas perturbed faster 
solitons appear to blow up.  We again consider initial data of 
the form (\ref{pertur}), this time with $c=-10$ and $\alpha=0.1$. We 
use $N_{x}=N_{y}=2^{11}$ Fourier modes and $N_{t}=50000$ time steps. At 
$t\sim 1.47$, the  relative energy conservation $H$ is no longer 
better than  $10^{-3}$ and the results for larger times are ignored. The 
solution at a later time is shown in Fig.~\ref{VNsolE0cm10}. As can 
be seen in the same figure on the right, there is a lack of resolution 
in the Fourier domain at this time.

\begin{figure}
\centering
    \includegraphics[width=0.45\textwidth]{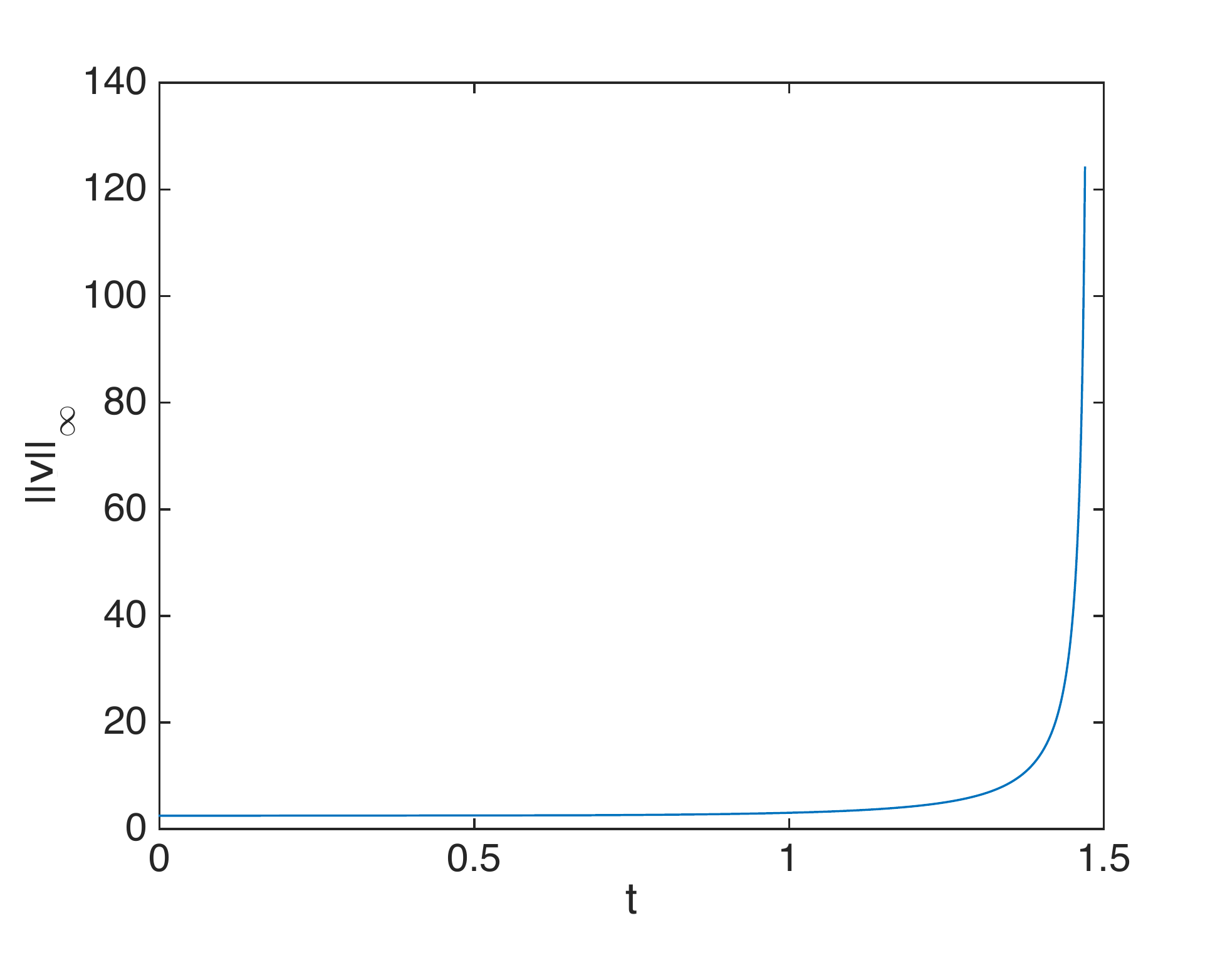}
    \includegraphics[width=0.45\textwidth]{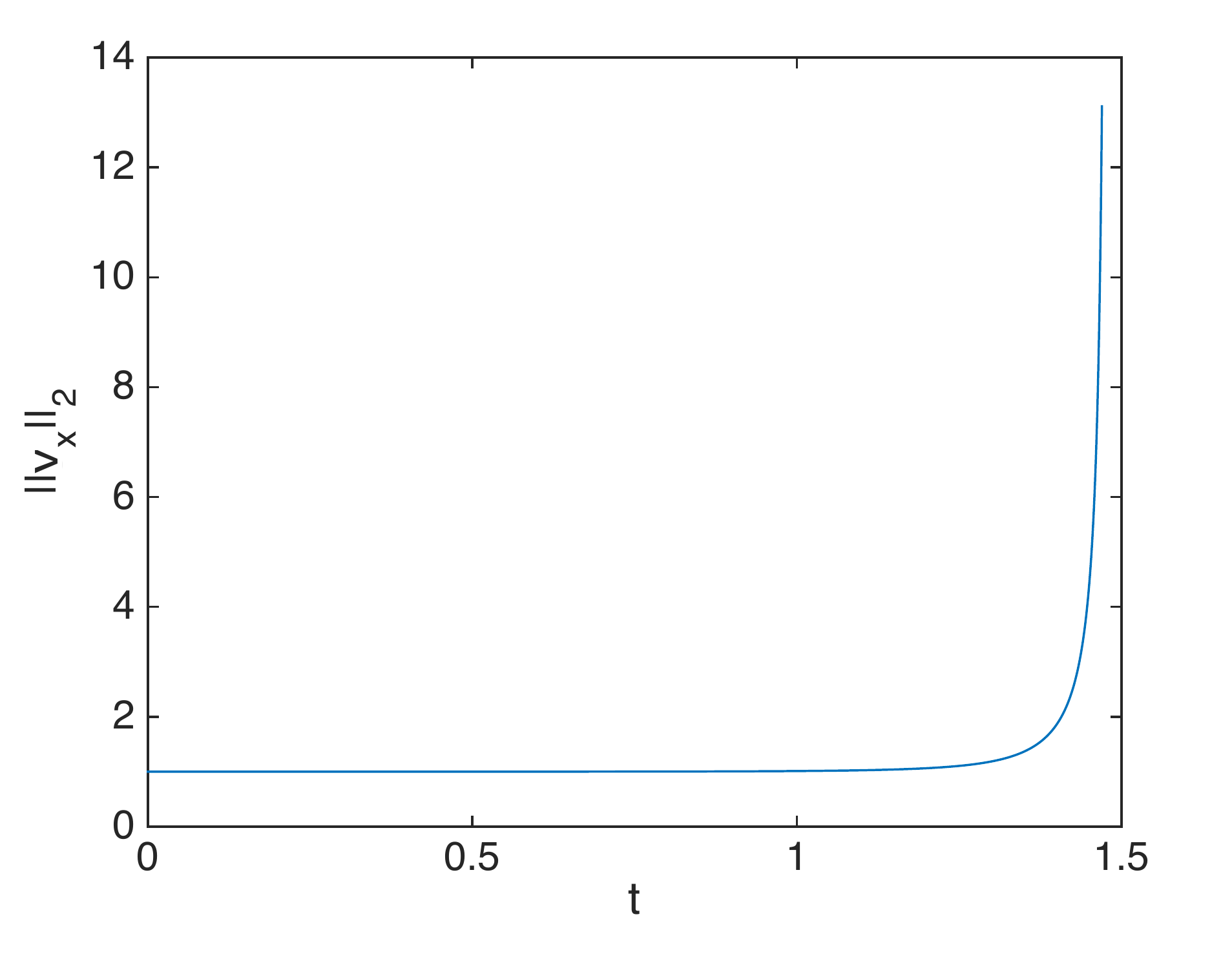}
 \caption{$L^{\infty}$ norm of the solution to the NV equation (\ref{NVgal}) 
 with $E=0$ for the initial data (\ref{pertur}) with $c=-10$ and 
 $\alpha=0.1$ in dependence on time on the left and the $L^{2}$ norm 
 of $v_{x}$  on the right. }
 \label{VNsolE0cm10norm}
\end{figure}

Both the $L^{\infty}$ norm and the $L^{2}$ norm of $v_{x}$ appear to 
explode as can be seen in Fig.~\ref{VNsolE0cm10norm}. But in contrast 
to the case $E=1$
there does not appear to be `metastable' structure at some 
intermediate time as in Fig.~\ref{VNsolE1cm20norm}. 

\begin{figure}
\centering
    \includegraphics[width=0.45\textwidth]{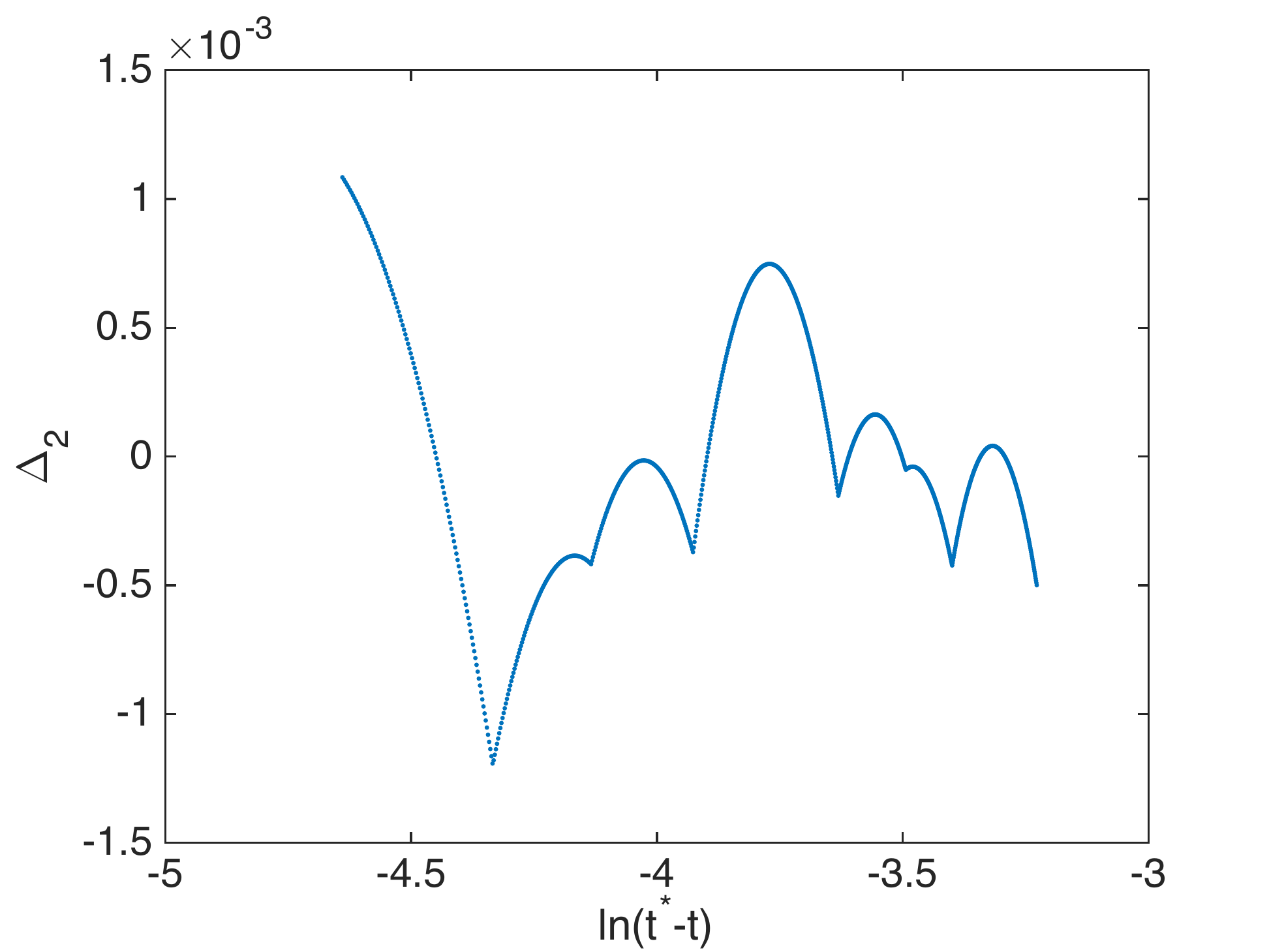}
    \includegraphics[width=0.45\textwidth]{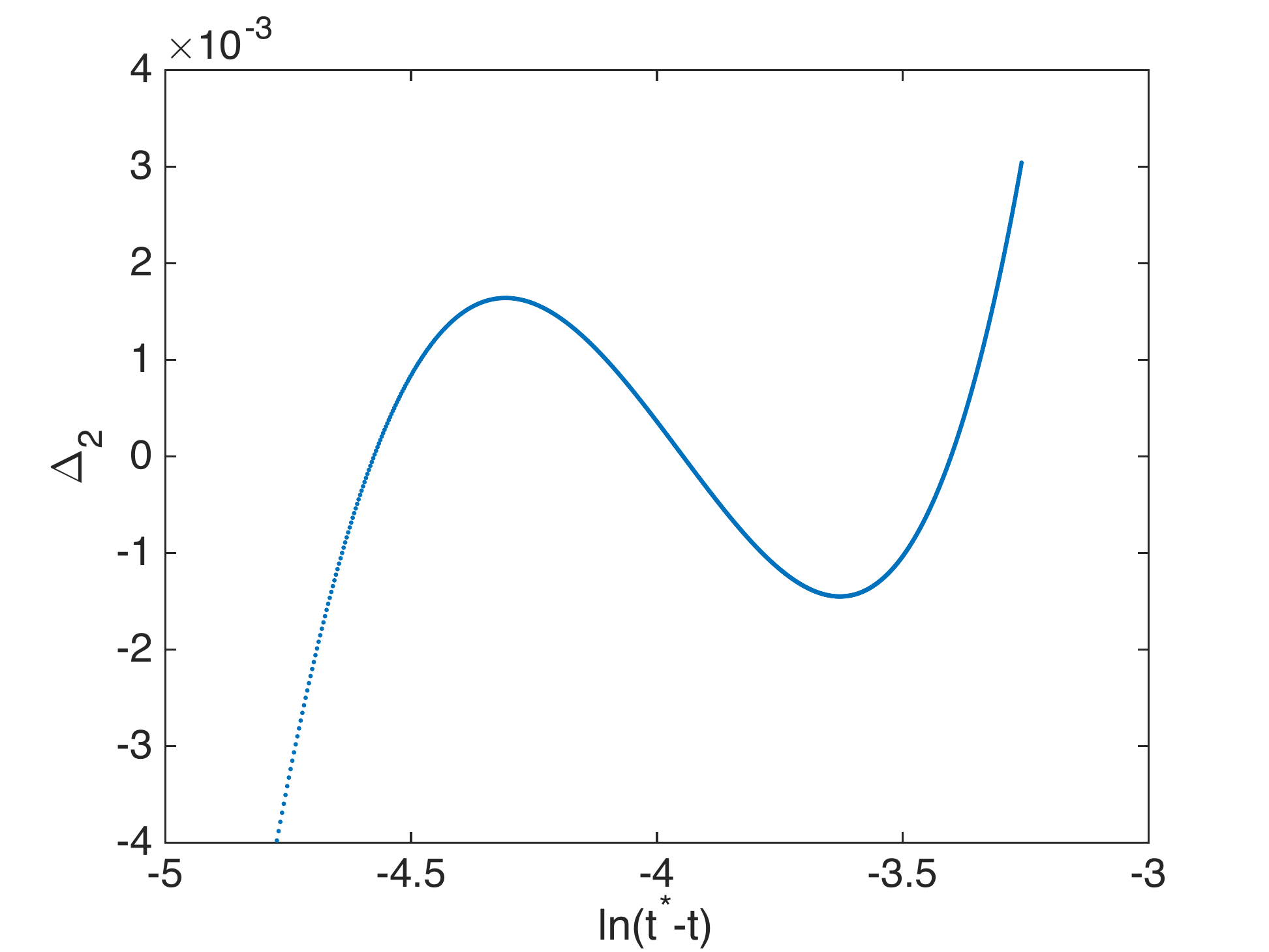}
 \caption{Fit of the norms in Fig.~\ref{VNsolE0cm10norm} to 
 $\gamma\ln (t^{*}-t)+\delta$; the quantity $\Delta_{2}:=|\ln 
 C-\gamma\ln (t^{*}-t)-\delta|$ for $C$ the $L^{\infty}$ norm of $ v$ on the 
 left and for $C$ the $L^{2}$ norm of $v_{x}$ on the right.}
 \label{VNsolE0cm10fit}
\end{figure}

We again  fit various norms for the last 1000 time steps to
$\gamma \ln (t^{*}-t)+\delta$ for constant $\gamma$ 
and $\delta$.  For the 
example in Fig.~\ref{VNsolE0cm10} we get for the norm 
$||v||_{\infty}$ the values $t^{*}=1.4797$, $\gamma=-1.03$ and 
$\delta=0.05$, and for the norm $||v_{x}||_{2}$ 
the values $t^{*}=1.4785$, $\gamma=-0.91$ and $\delta=-1.77$. The 
quality of the fittings can be seen in Fig.~\ref{VNsolE0cm10fit}, the 
fitting errors are of the order of $10^{-3}$. The results 
do not change much if the fitting is done for the last 500 time 
steps. The compatibility between the found blow-up times shows the 
consistency of the fitting. The results indicate once more a 
 blow-up of the type   
(\ref{eq:Crit_Lt}) with $\gamma_{1}=-1$. 

\begin{figure}
\centering
    \includegraphics[width=0.45\textwidth]{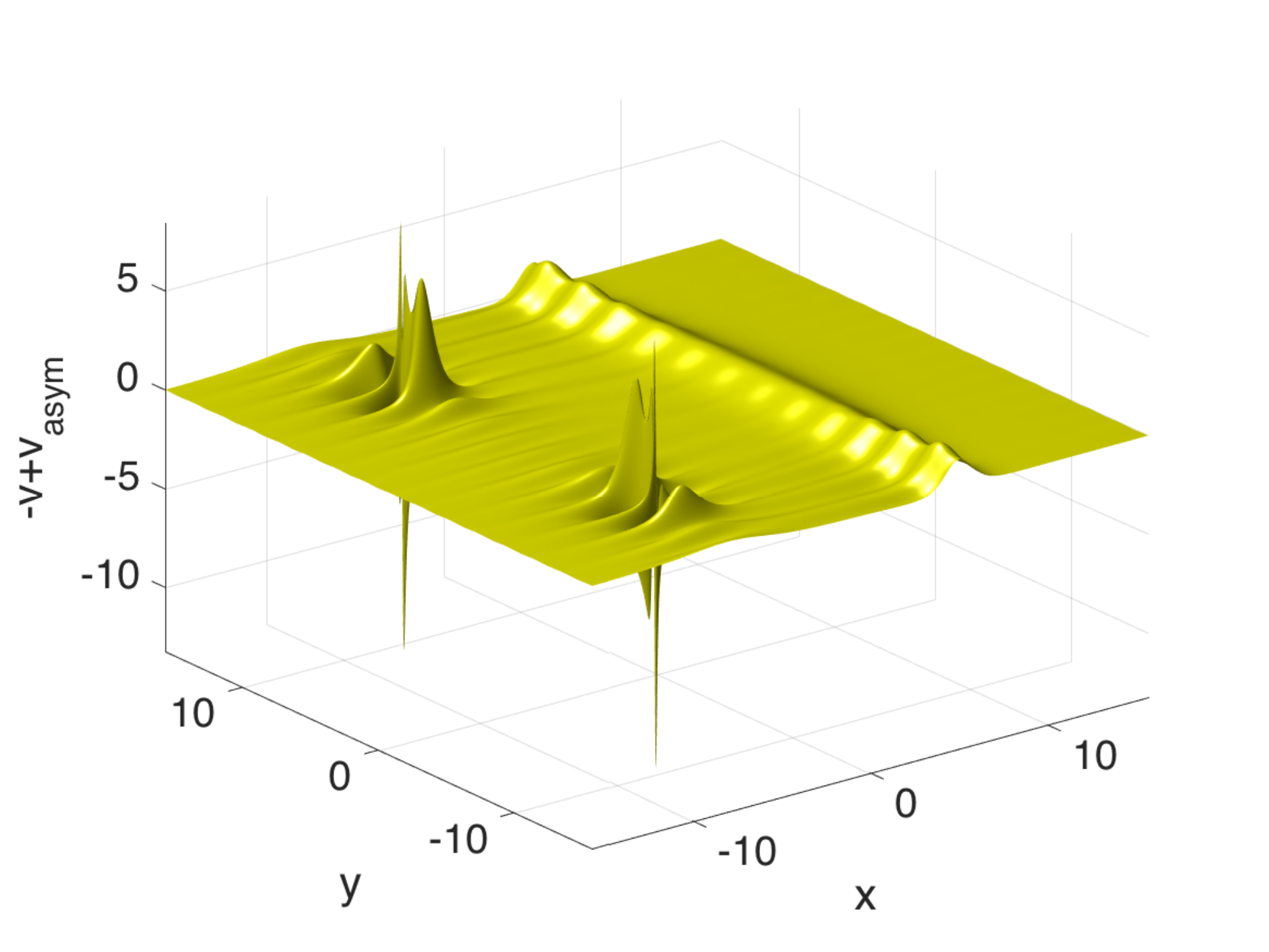}
    \includegraphics[width=0.45\textwidth]{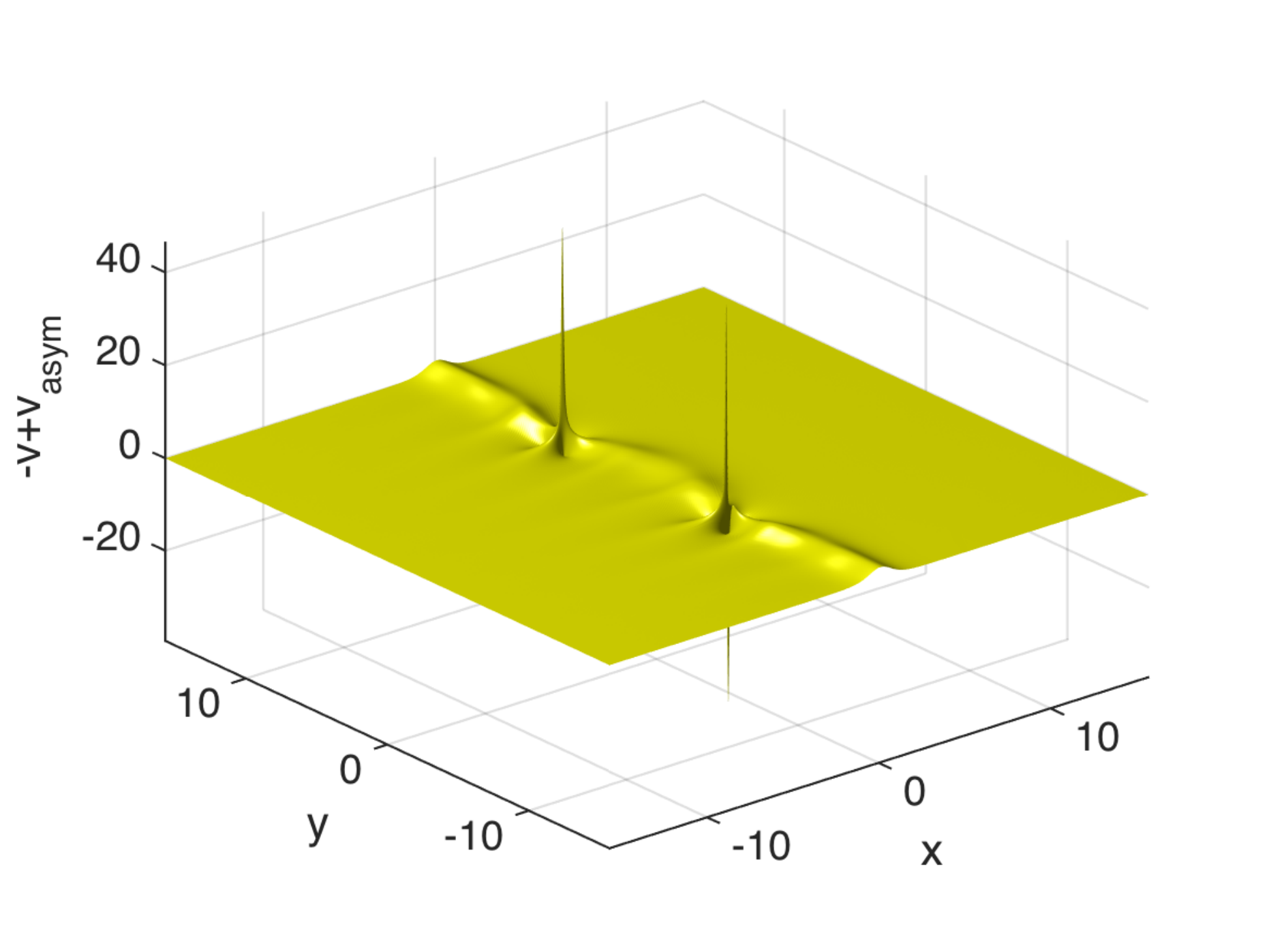}
 \caption{Difference between the numerical solution and a rescaled 
 lump (\ref{new_lumps}) with $a=b=0$ localized at the apparent 
 blow-ups and rescaled according to (\ref{gKP4}); on the left for the 
 situation for   Fig.~\ref{VNsolE1cm20} and on the right 
 for Fig.~\ref{VNsolE0cm10}. }
 \label{NVsolfit}
\end{figure}

The blow-up mechanism that appears under the situations in 
Fig.~\ref{VNsolE1cm20} and Fig.~\ref{VNsolE0cm10} also implies that the profile of the self 
similar blow-up is given by a travelling wave solution of the NV 
equation for $E=0$, i.e. equation (\ref{inf}) with $a^{\infty}=0$. 
For $E=0$, two families of localized travelling wave solutions are known, the 
lumps (\ref{Chsol}) (or, more generally, (\ref{tai}) with $c=0$) and (\ref{new_lumps}). Since the former vanish at 
the origin, their center of symmetry, they cannot be a candidate for a 
blow-up profile of the type observed here, blow-ups in isolated 
points. The lumps (\ref{new_lumps}) are the only known candidates 
with the wanted behaviour. We choose $a=b=0$ (this just corresponds to a 
shift and rescaling of the solution). We determine the localization and value of 
the minima of the solution near the two peaks in 
Fig.~\ref{VNsolE1cm20} and Fig.~\ref{VNsolE0cm10}. The quantity $L$ in (\ref{gKP4}) is 
fixed by fitting the rescaled soliton to the respective minima. 
 In Fig.~\ref{NVsolfit} we show the difference 
between the solution in Fig.~\ref{VNsolE1cm20} and the lump 
(\ref{new_lumps}) rescaled according to (\ref{gKP4}) at each minimum 
on the left, and the corresponding figure for the situation of 
Fig.~\ref{VNsolE0cm10} on the right. 
It can be seen that these rescaled lumps catch the main profile. This 
is done in a much better way on the left than on the right, which 
indicates that  the blow-up profile is indeed given by a dynamically 
rescaled lump (\ref{new_lumps}). In the second case, it appears that we are
not close enough to the 
blow-up to have an even better agreement since we ran out of 
resolution too early.  

Note that we observe qualitatively the same behavior for $E=-1$ as 
for $E=0$ and $E=1$: perturbations of small enough KdV solitons are 
just radiated away and the soliton is thus stable. Larger solitons 
appear to be unstable against a blow-up. Since the results are very 
similar to what is shown above, we do not give details for this case. 

\section{Localized initial data}
\label{localized}
In this section we study localized initial data of the form 
\begin{equation}
    v(x,y,0) = \beta \exp(-x^{2}-y^{2})
    \label{initial},
\end{equation}
where $\beta$ is some real constant. Note that these initial data 
have finite mass in contrast to the KdV soliton treated in the 
previous section. Though we work numerically on $\mathbb{T}^{2}$ 
instead of $\mathbb{R}^{2}$ and thus always with finite masses, this 
different setting leads to some qualitative differences with respect 
to what was observed in the context of the KdV soliton.

\subsection{The KPI limit: $E\gg1$}
\begin{figure}
\centering
    \includegraphics[width=0.45\textwidth]{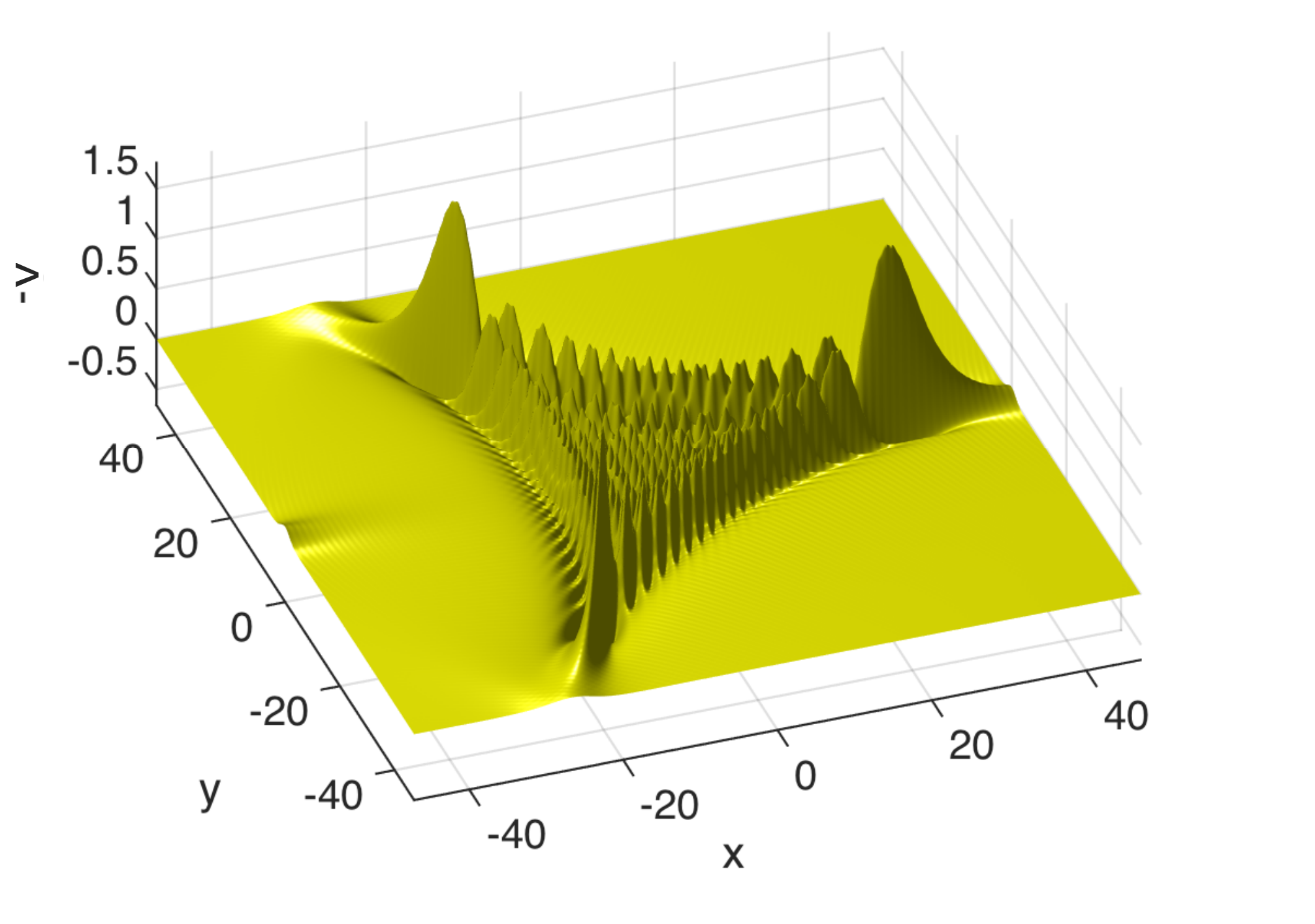}
    \includegraphics[width=0.45\textwidth]{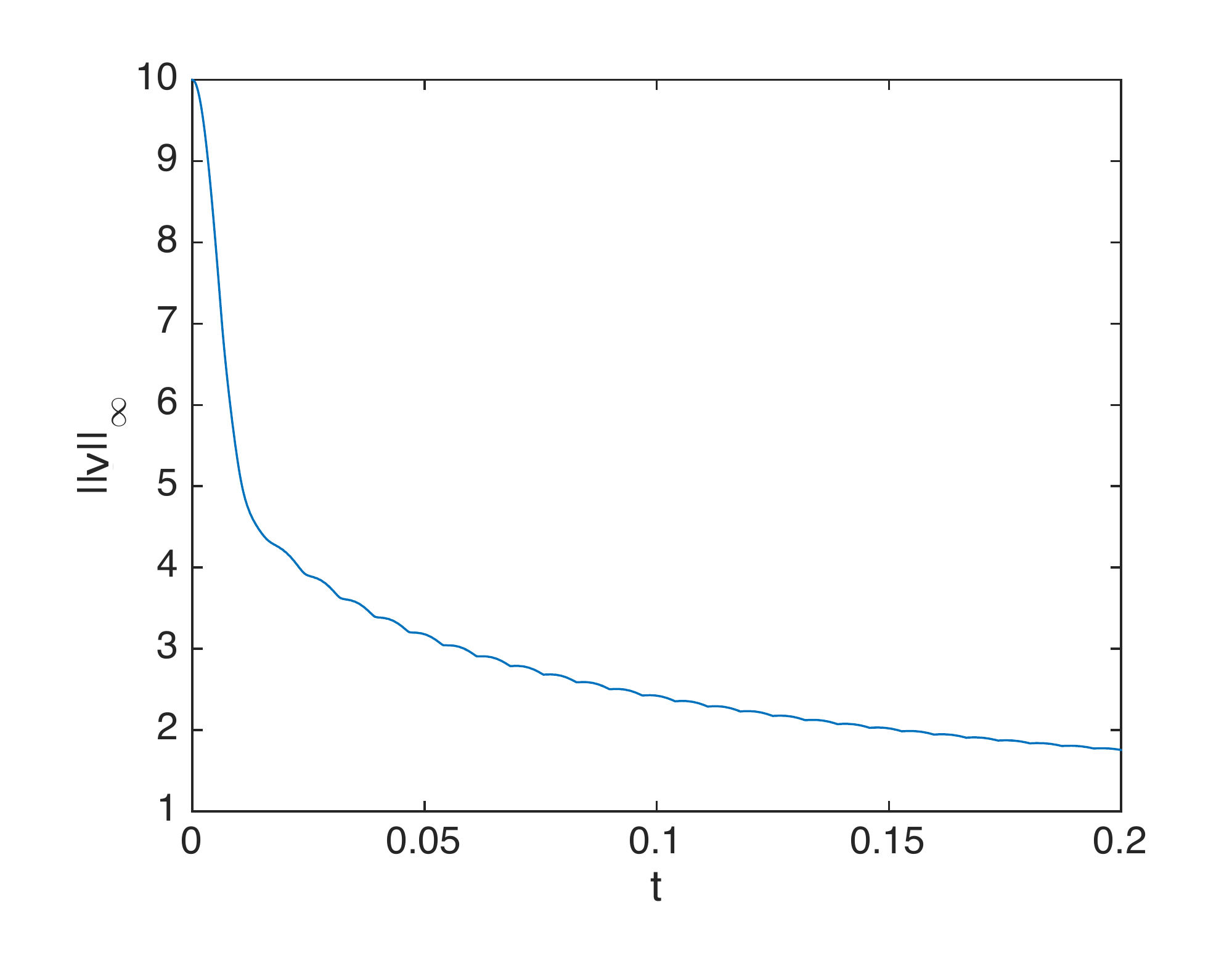}
 \caption{Solution of the NV equation (\ref{NV_eq}) with $E=10$ for the initial 
 data (\ref{initial}) and $\beta=-10$; on the left the solution 
for $t=0.2$, on the right the $L^{\infty}$ norm of the solution in 
dependence of time. }
 \label{NVE10m10gauss}
\end{figure}

We first consider the case $E=10$ for which NV  should 
approach in some sense KP I. We use $N_{x}=N_{y}=2^{10}$ modes for 
$x,y\in15[-\pi,\pi]$ and $N_{t}=1000$ time steps for $t\leq 0.2$. It 
can be seen in Fig.\ref{NVE10m10gauss} that the initial pulse with 
rotational symmetry is dispersed to infinity along three lines with an angle of 
120 degrees between them. Note that the figures are very 
similar for positive values of $\beta$ in (\ref{initial}). 

The $L^{\infty}$ norm of the 
solution in the same figure indicates that  the initial pulse is 
just dispersed on the studied time scales. This does not change if we 
run this example on longer scales of time and space. Thus it appears 
the initial data in this example are just dispersed to infinity 
without formation of a lump. We cannot exclude of course that a lump 
will appear at later times. But calculations for KPI in 
\cite{KR14,DGK15} show that lumps are difficult to observe in the 
context of localized initial data: there initial data 
with support on length scales of order $1/\epsilon$ were considered 
on time scales of the order $1/\epsilon$ with $\epsilon\ll 1$. In 
that case \emph{dispersive shock waves} were observed, i.e., zones of 
rapid modulated oscillations in the solution. It seems 
that lumps can only appear in these examples for large times compared to the time where 
the dispersive shock forms. To access this in a reliable way, 
stronger computers would be needed than the ones used for the present 
work. 

Thus it is not surprising that we do not see lumps in the solutions 
for the NV equation for initial data of the form (\ref{initial}) on 
the studied time scales. This does not change if we take even larger 
values of $E$ or larger values of $|\beta|$. Note, however, that the 
transition between the regime $E\approx 0$, where blow-up is possible, 
and the regime for large $E$, where a KPI type behavior is expected, 
is smooth. For very large values of $\beta$ ($\beta\sim-100$), 
there appears to be a blow-up. 

\subsection{The KPII limit: $E\ll -1$}
\begin{figure}
\centering
    \includegraphics[width=0.45\textwidth]{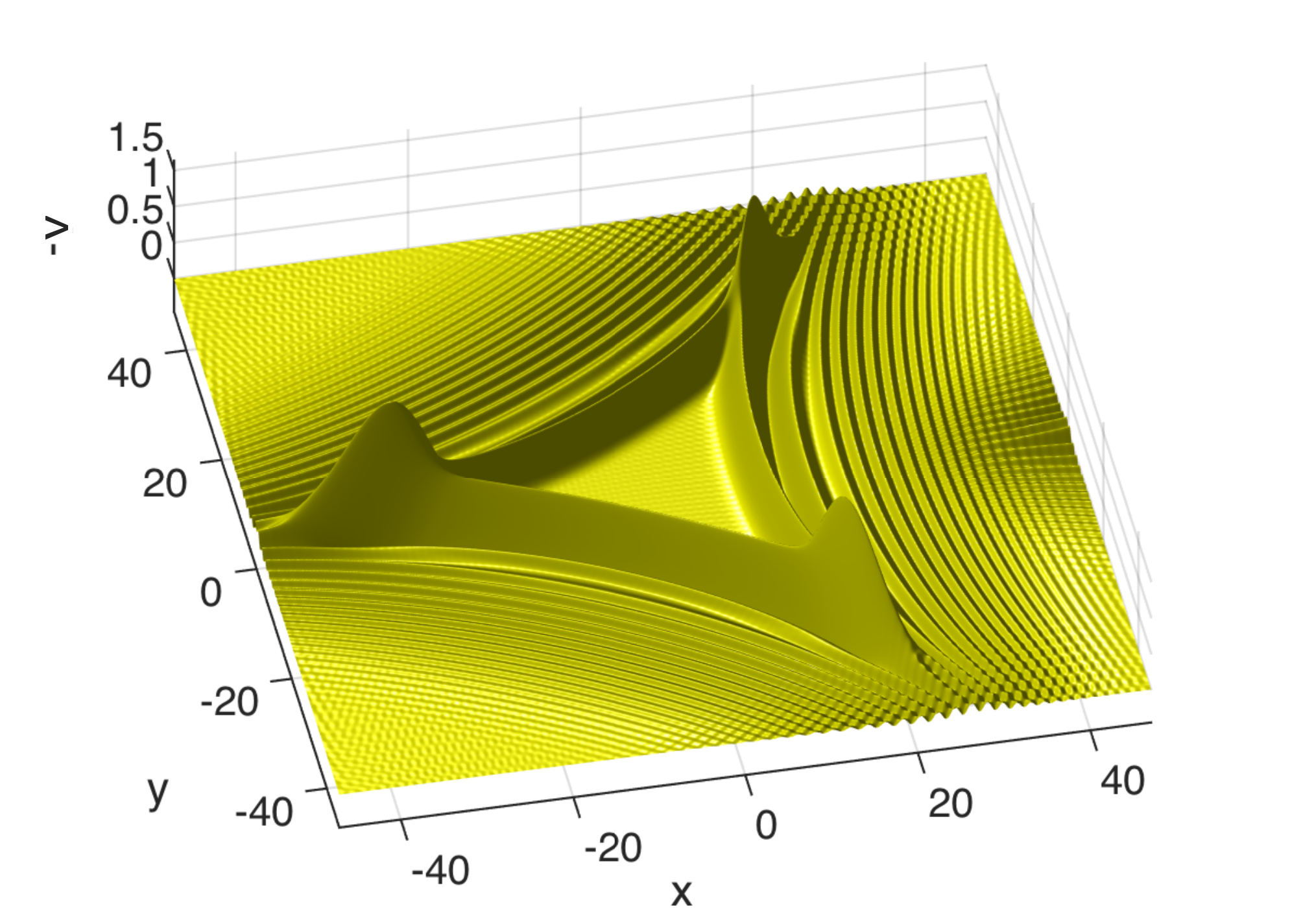}
    \includegraphics[width=0.45\textwidth]{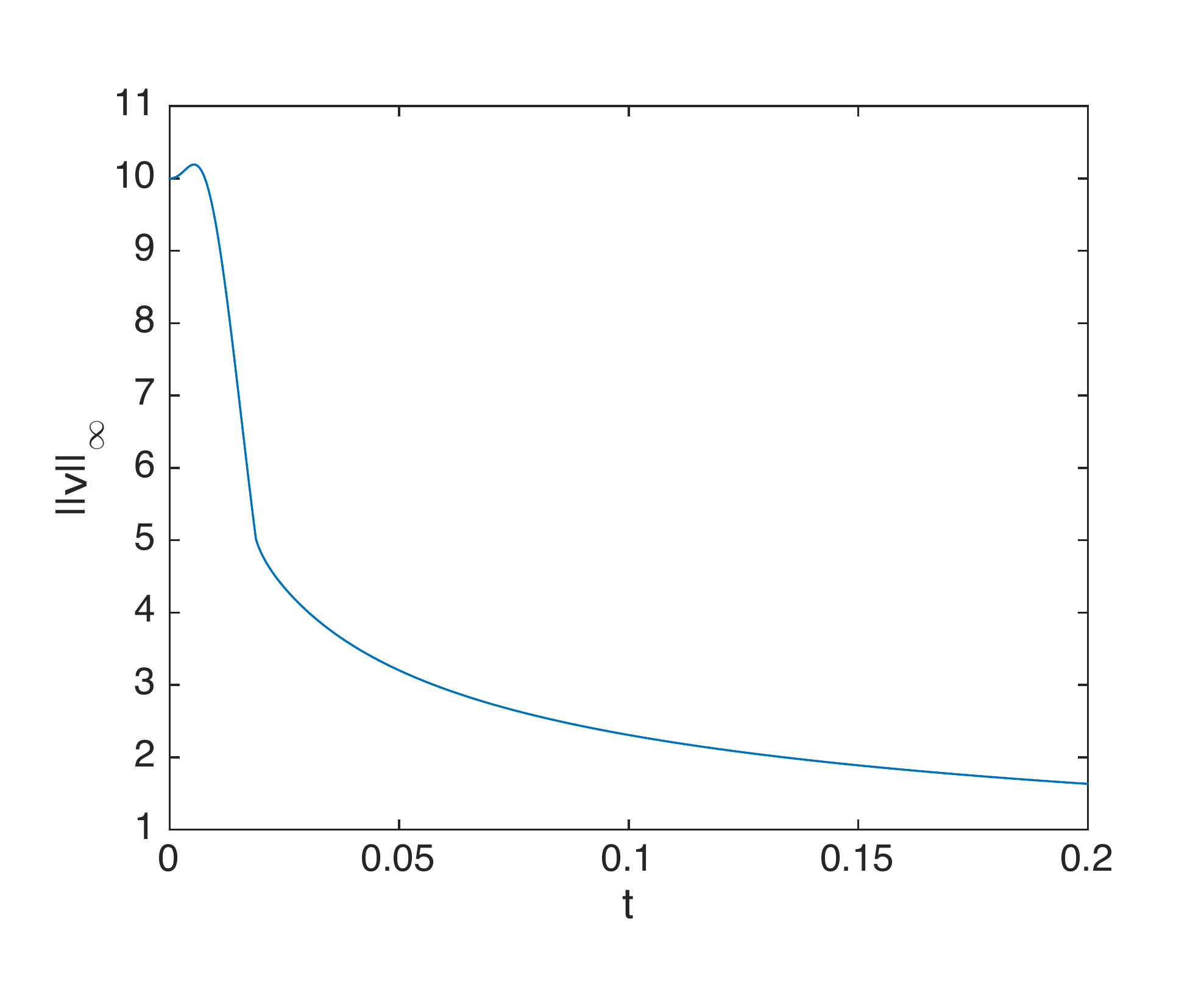}
 \caption{Solution of the NV equation (\ref{NV_eq}) with $ E =-10 $ for the initial 
 data (\ref{initial}) and $\beta=-10$; on the left the solution 
for $t=0.2$, on the right the $L^{\infty}$ norm of the solution in 
dependence of time. }
 \label{NVEm10m10gauss}
\end{figure}

For large negative values of $E$, the NV equation approaches the 
KPII equation. We show the solution for $E=-10$ for the initial data 
(\ref{initial}) with  $\beta=-10$ in Fig.~\ref{NVEm10m10gauss}. It 
can be seen that the initial pulse is just radiated away to infinity. 
Again the 120 degree symmetry is observed for rotationally symmetric 
initial data. No stable structures appear in the evolution as can be 
seen also from the $L^{\infty}$ norm of the solution in the same 
figure.  Since KPII has no lump solutions, and since its solutions 
are expected to exist globally in time, this is the expected behavior 
for this regime of NV. Note that in contrast to the same 
initial data for $E=10$ in Fig.~\ref{NVE10m10gauss}, there are no 
oscillations orthogonal to the propagation direction of the 
pulses. The latter could lead to lumps if sufficient mass is present 
for $E\gg 1$, but such structures should not appear in the limit 
$-E\gg1$. Instead the oscillations are here tangential to the 
triangular structure into which the initial pulse evolves. 

\begin{figure}
\centering
    \includegraphics[width=0.7\textwidth]{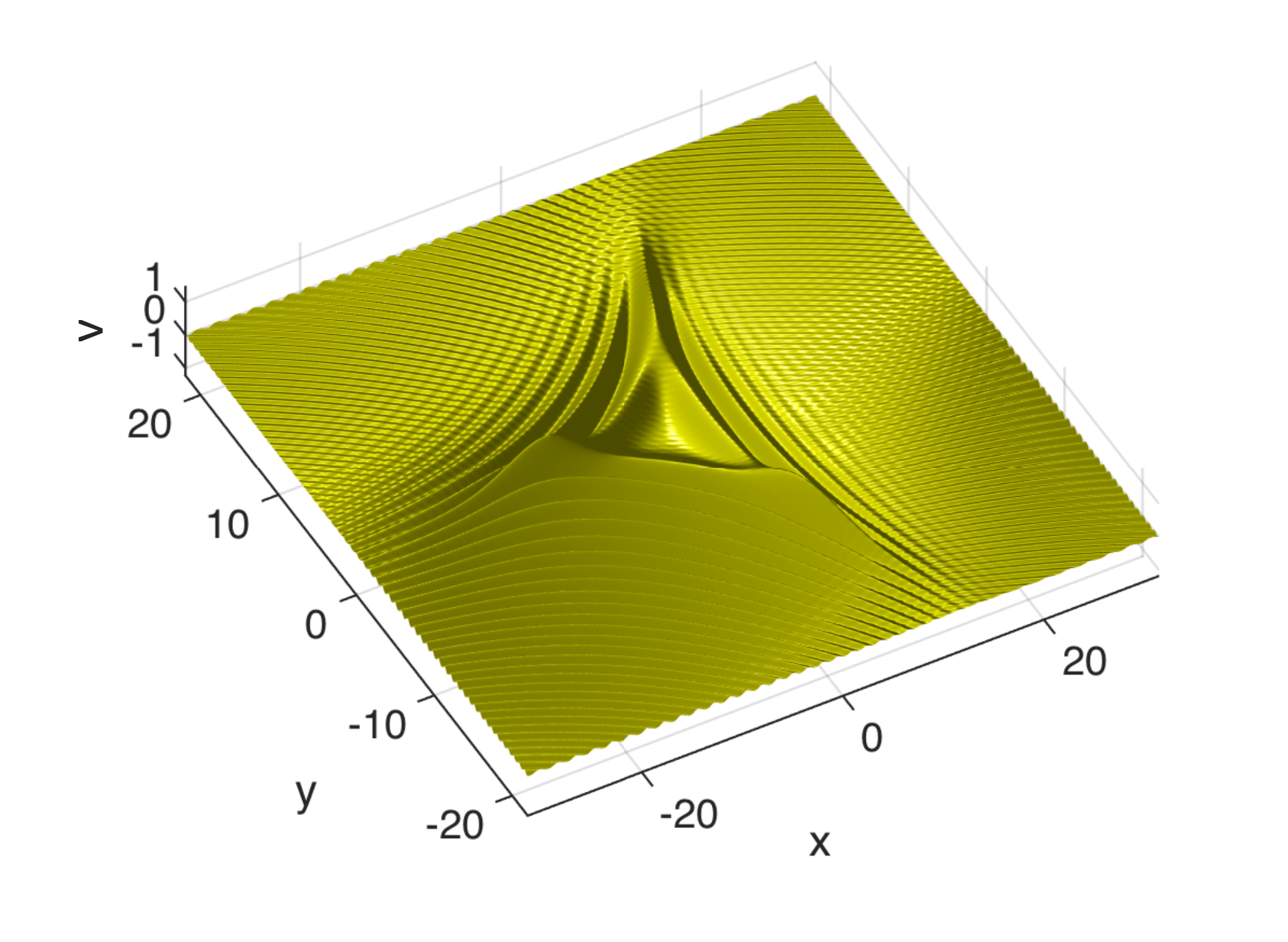}
 \caption{Solution of the NV equation (\ref{NV_eq}) for the initial 
 data $v(x,y,0)=\exp(-x^{2}-5y^{2}-3xy)$ and $E=-10$ for $t=0.06$. }
 \label{NV10asymEm10t006}
\end{figure}

If we take initial data without rotational symmetry, the solution 
will not have this 120 degree symmetry, but its propagation will 
follow the 120 degree pattern. This can be seen in 
Fig.~\ref{NV10asymEm10t006} for the initial data 
$v(x,y,0)=\exp(-x^{2}-5y^{2}-3xy)$ which are not rotationally 
symmetric even for $|x|,|y|\to\infty$. The computation is carried out 
for $x\in 10[-\pi,\pi]$ and $y\in7[-\pi,\pi]$ with 
$N_{x}=N_{y}=2^{10}$ Fourier modes and $N_{t}=10^{3}$ time steps for 
$t\leq 0.06$. 

\subsection{Intermediate values of $E$}
\begin{figure}
\centering
    \includegraphics[width=0.45\textwidth]{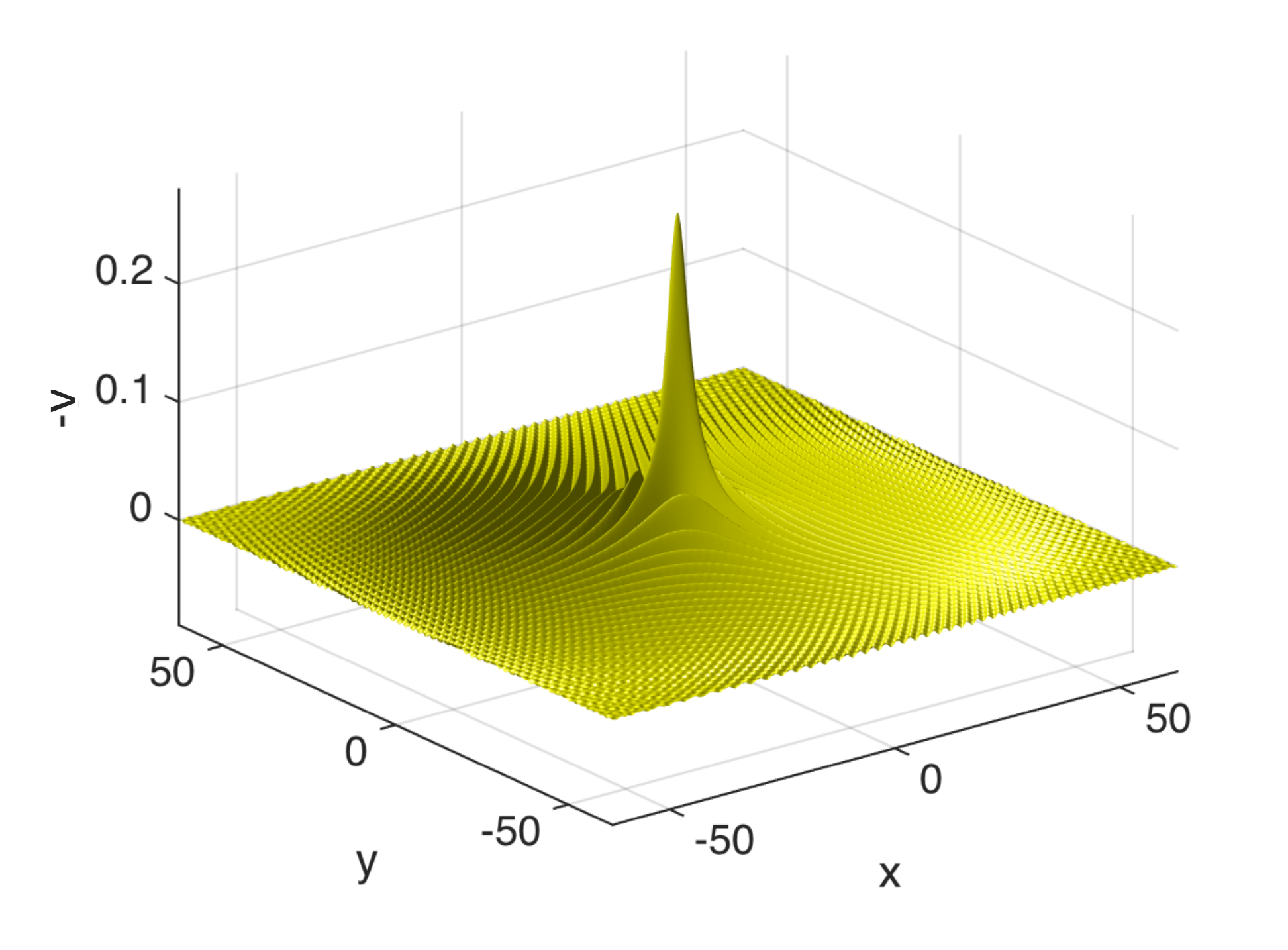}
    \includegraphics[width=0.45\textwidth]{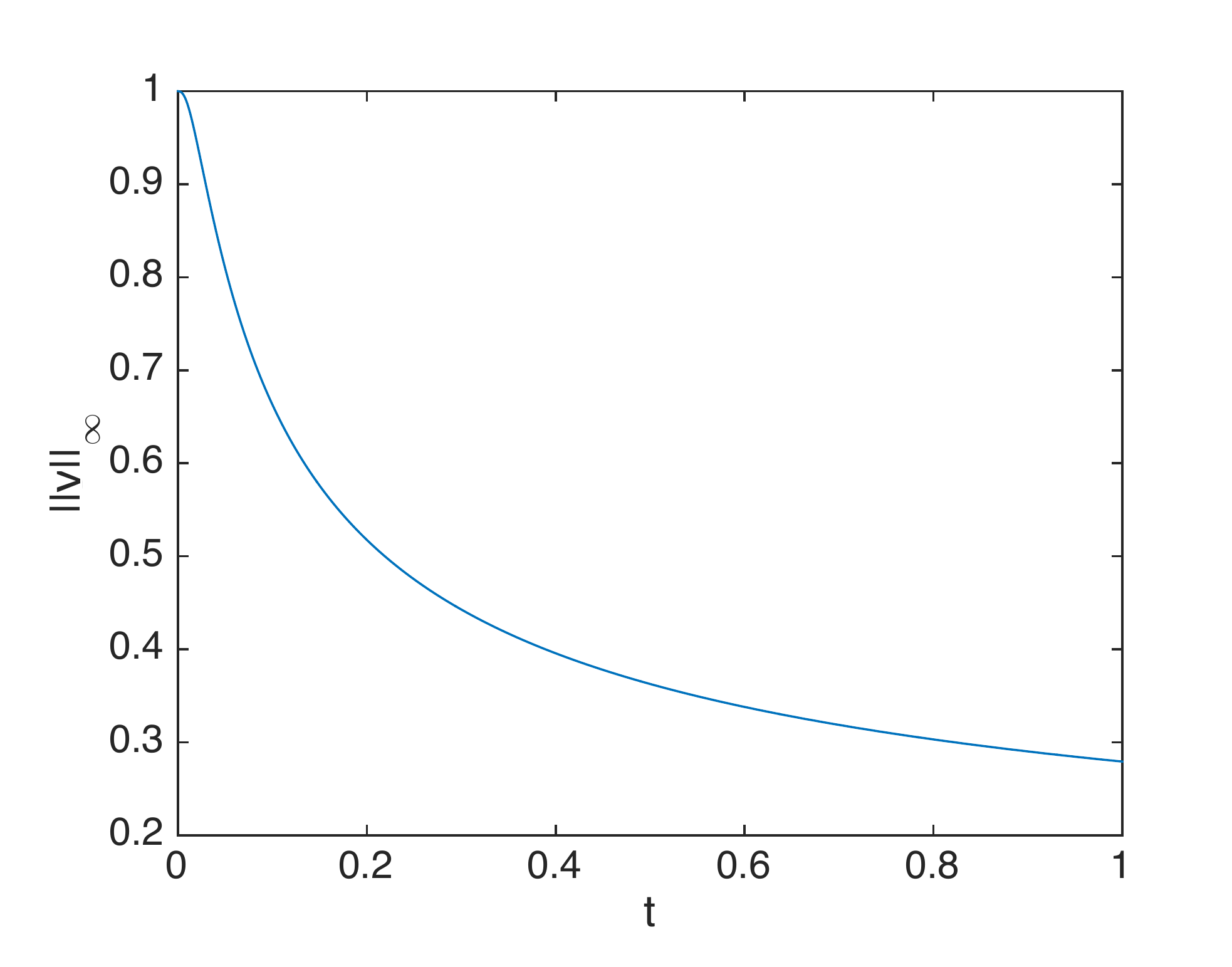}
 \caption{Solution of the NV equation (\ref{NV_eq}) for the initial 
 data (\ref{initial}) and $\beta=-1$ for $E=0$; on the left the solution 
for $t=1$, on the right the $L^{\infty}$ norm of the solution in 
dependence of time. }
 \label{NVE0mgauss}
\end{figure}

For $E=0$, the solution to the NV equation for the initial data 
(\ref{initial}) with $\beta=-1$ appears to be global in time. The 
computation is done with $N_{x}=N_{y}=2^{10}$ Fourier modes for 
$[x,y]\in20[-\pi,\pi]^{2}$. The 
resulting solution can be seen in Fig.~\ref{NVE0mgauss}. It appears 
to be just dispersed away towards infinity along the 120 degree 
lines. This is confirmed by the $L^{\infty}$ norm of the solution on 
the right of Fig.~\ref{NVE0mgauss} which appears to decrease 
monotonically. 

\begin{figure}
\centering
    \includegraphics[width=0.45\textwidth]{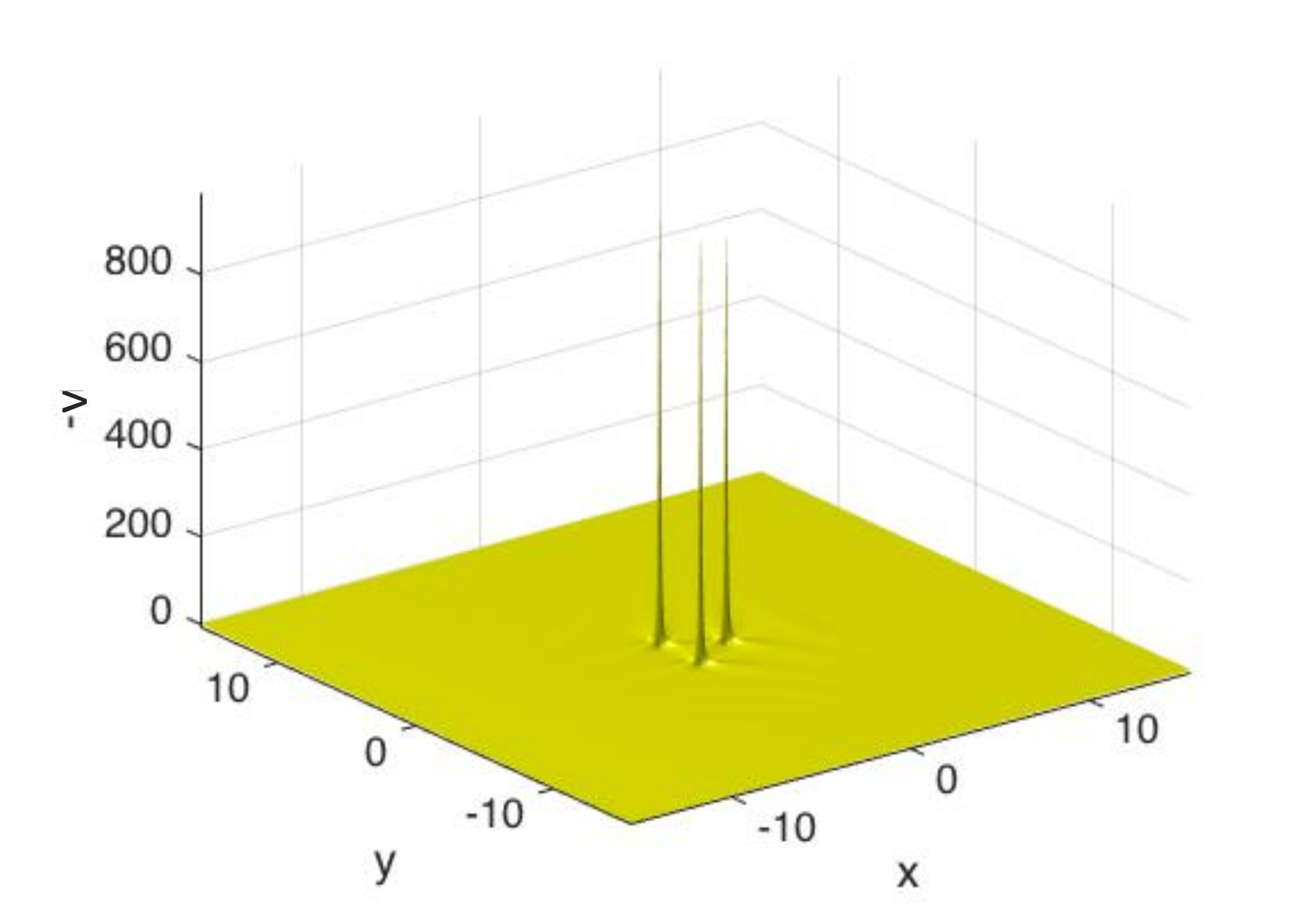}
    \includegraphics[width=0.45\textwidth]{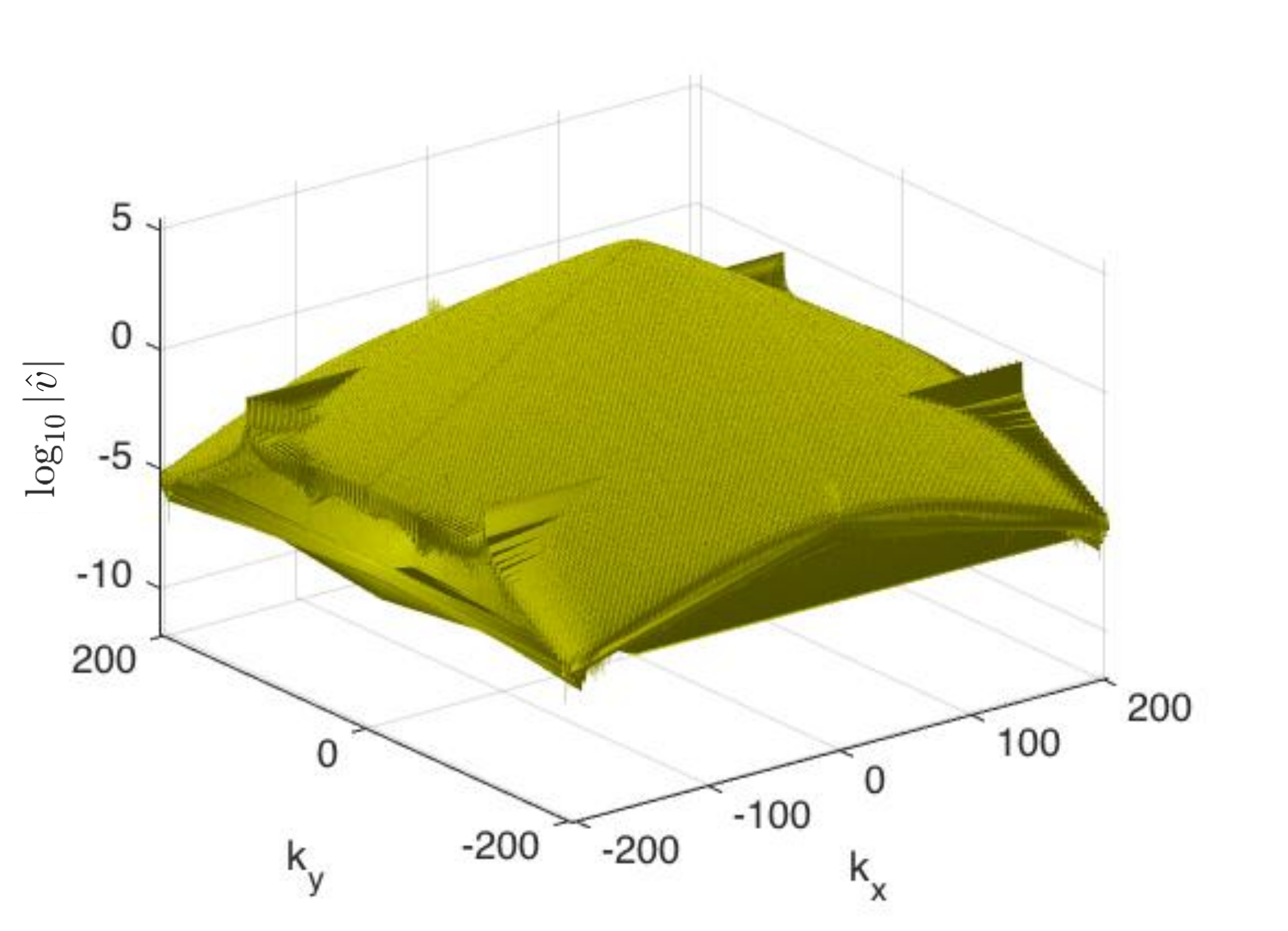}
 \caption{Solution of the NV equation (\ref{NVgal}) for the initial 
 data (\ref{initial}) and $\beta=-10$ for $E=0$; on the left the solution 
for $t=0.1365$ close to a blow-up, on the right the modulus of the 
Fourier coefficients at the same time. }
 \label{NVm10gaussE0}
\end{figure}

However, for $E=0$ and the initial data (\ref{initial}) with $\beta=-10$, 
there appears to be a blow-up in finite time. The code is run for 
$[x,y]\in5[-\pi,\pi]\times[-\pi,\pi]$ with $N_{x}=N_{y}=2^{11}$ 
Fourier modes and $N_{t}=10^{5}$ time steps for $t\leq 0.14$. The 
code is stopped at $t=0.1365$ where the relative conservation of energy $H$ drops below $10^{-3}$. The solution at this time 
is shown in Fig.\ref{NVm10gaussE0}. It can be seen that,
because of the rotational symmetry of the initial data, the solution  shows the 120 
degree symmetry. Thus the blow-up appears to happen in 3 points at 
the same time. The Fourier coefficients at the final time on the 
right of Fig.~\ref{NVm10gaussE0} indicate once more that resolution 
in $t$ appears to lack in these blow-ups before this happens in the 
spatial coordinates. 

\begin{figure}
\centering
    \includegraphics[width=0.45\textwidth]{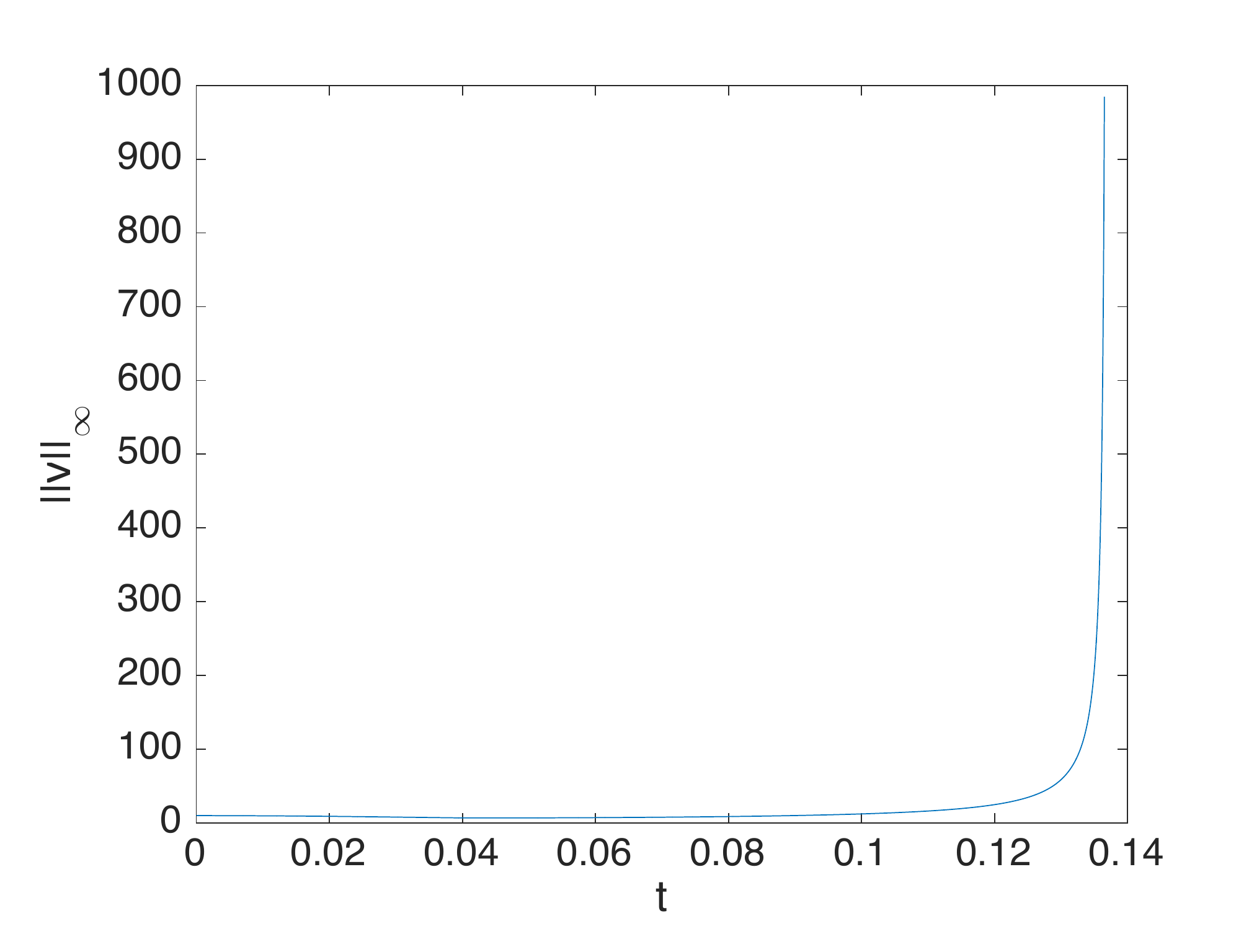}
    \includegraphics[width=0.45\textwidth]{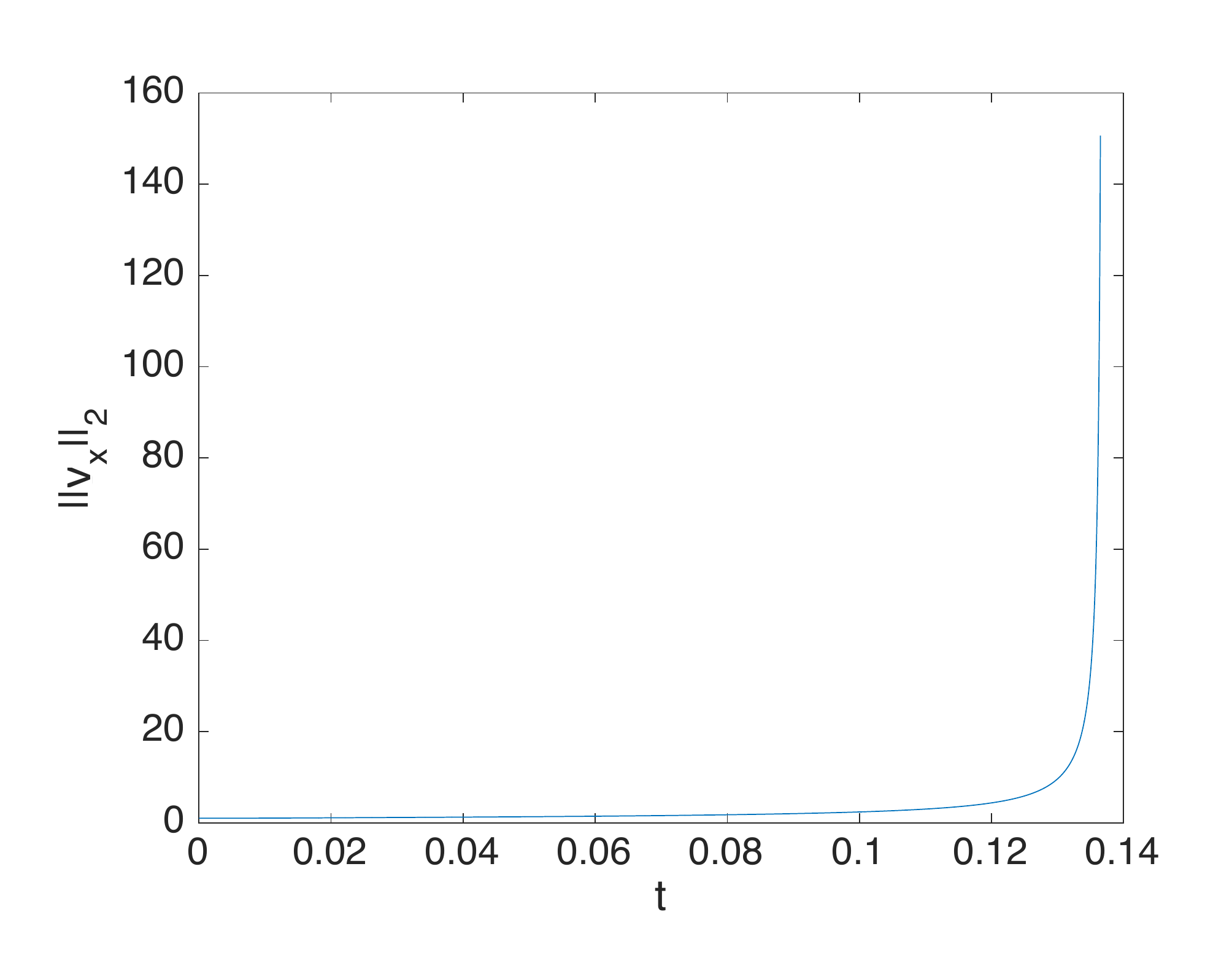}
 \caption{$L^{\infty}$ norm of the solution to the NV equation (\ref{NVgal}) 
 with $E=0$ for the initial data (\ref{initial}) with  
 $\beta=-10$ in dependence on time on the left and the $L^{2}$ norm 
 of $v_{x}$  on the right. }
 \label{NVE0m10gaussnorm}
\end{figure}

Once more both the $L^{\infty}$ norm of $v$ and the $L^{2}$ norm of $v_{x}$ 
indicate a blow-up as can be seen in Fig.~\ref{NVE0m10gaussnorm}. 
        
\begin{figure}       
\centering
    \includegraphics[width=0.45\textwidth]{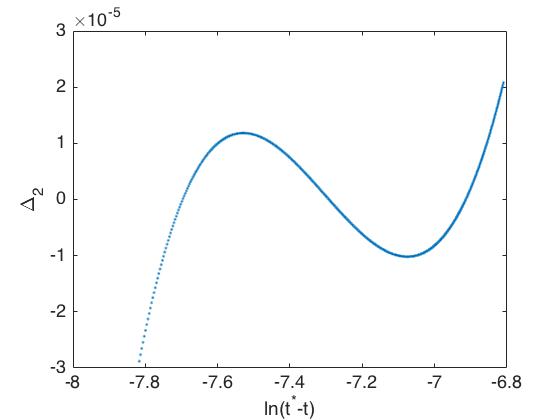}
    \includegraphics[width=0.45\textwidth]{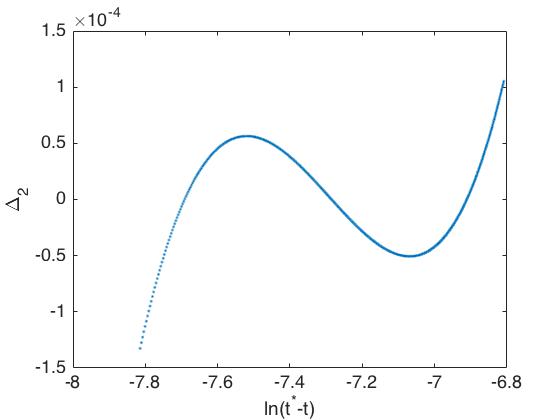}
 \caption{Fit of the norms in Fig.~\ref{NVE0m10gaussnorm} to 
 $\gamma\ln (t^{*}-t)+\delta$; the quantity $\Delta_{2}:=|\ln 
 C-\gamma\ln (t^{*}-t)-\delta|$ for $C$ the $L^{\infty}$ norm on the 
 left and for $C$ the $L^{2}$ norm of $v_{x}$ on the right. }
 \label{VNm10gaussE0fit}
\end{figure}

To understand the mechanism of a potential blow-up, we again perform a fit 
of  $\ln C$ to $\gamma \ln (t^{*}-t)+\delta$, where $C$ is either the 
norm $||v||_{\infty}$ or the norm $||v_{x}||_{2}$, $t^*$ is the blow-up time and $\gamma$ 
and $\delta$ are constants. This fit is performed for the last 500 
time steps (the results are very similar for the last 100 time steps)
with the optimization algorithm \cite{fminsearch}. For the 
example in Fig.~\ref{NVE0m10gaussnorm} we get for the norm 
$||v||_{\infty}$ the values       $t^{*}=0.1369$, $\gamma=-0.997$ and         
$\delta= -0.902$, and for the norm $||v_{x}||_{2}$                
the values $t^{*}=0.1369$, $\gamma=-0.983$ and $\delta=-2.667$. The 
quality of the fittings can be seen in Fig.~\ref{VNm10gaussE0fit}, the 
fitting errors are of the order of $10^{-4}$.  The compatibility between the found blow-up times shows the 
consistency of the fitting. The results again indicate that the factor $L$ 
in (\ref{gKP4})
should be proportional to $\sqrt{t^{*}-t}$,  as in 
(\ref{eq:Crit_Lt}) with $\gamma_{1}=-1$ which gives some indication that this is indeed 
the generic blow-up mechanism to be observed in NV solutions. 

\begin{figure}       
\centering
    \includegraphics[width=0.7\textwidth]{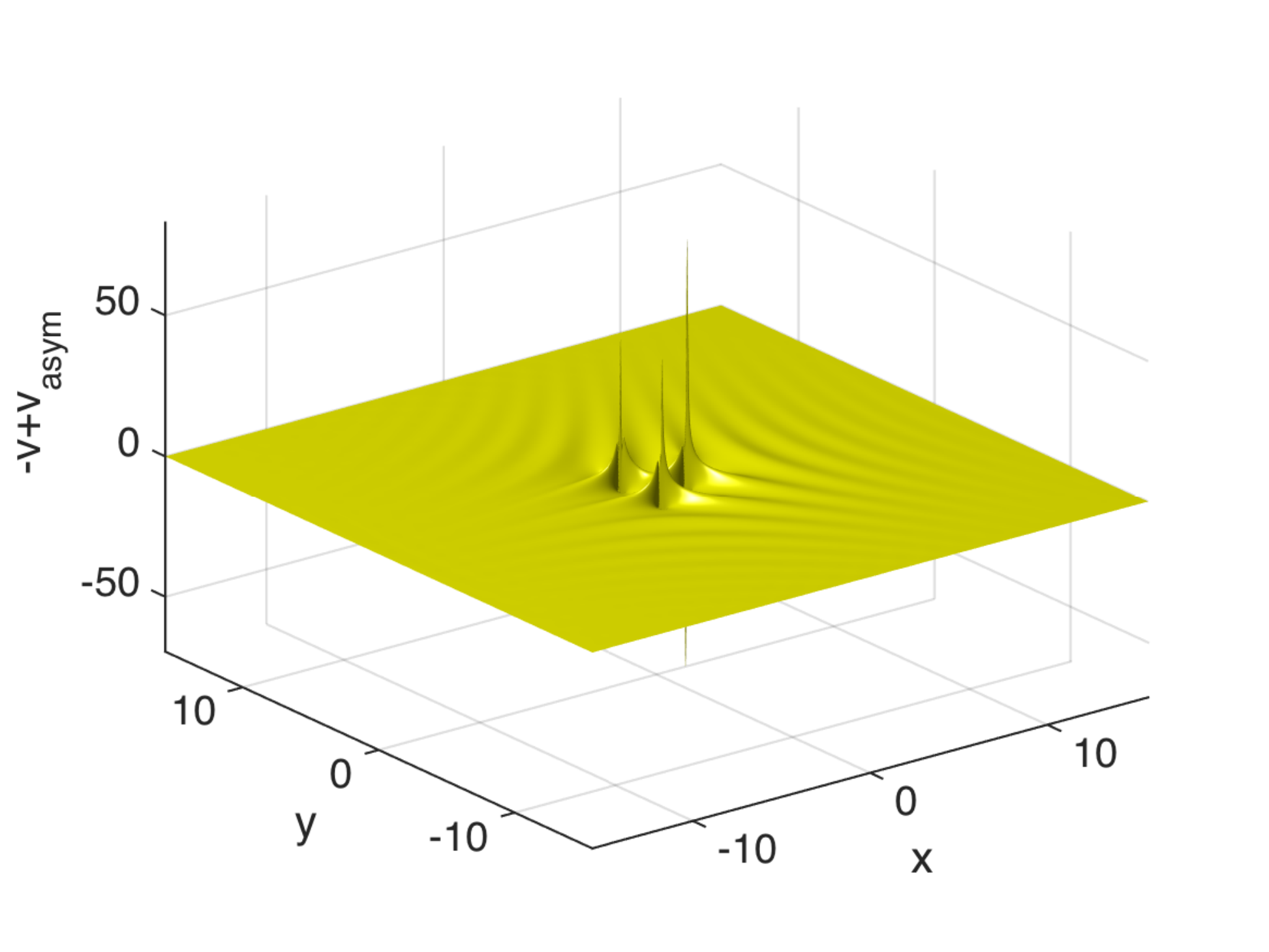}
 \caption{Difference of the NV solution  in Fig.~\ref{NVm10gaussE0} and 
 three lumps (\ref{new_lumps}) rescaled according to (\ref{gKP4}) at 
 each minimum.}
 \label{NVm10gaussE0solfit}
\end{figure}

Again this blow-up mechanism implies that the profile of the self 
similar blow-up is given by a travelling wave solution of the NV 
equation for $E=0$. 
We determine the localization and value of 
the minima of the solution near the three  peaks in 
Fig.~\ref{NVm10gaussE0}. The quantity $L$ in (\ref{gKP4}) is 
fixed by fitting the rescaled soliton to the respective minima. 
Note that the 120 degree symmetry is not exactly observed at the 
recorded time, thought it appears to be an attractor for the asymptotic 
state of the solution. The fitting, however, is performed for each 
point separately. In Fig.~\ref{NVm10gaussE0solfit} we show the difference 
between the solution in Fig.~\ref{NVm10gaussE0} and the lump 
(\ref{new_lumps}) rescaled according to (\ref{gKP4}) at each minimum. 
It can be seen that these rescaled lumps catch the main profile (the 
difference is smaller than the solution by more than an order of 
magnitude), but that we are, as expected, not close enough to the 
blow-up to have an even better agreement.  

Note that very similar behavior is observed for $E=\pm1$ as for $E=0$ 
which is why we do not discuss these cases in more detail: initial 
data of sufficiently small $L^{2}$ norm are just radiated away, 
whereas initial data of large enough $L^{2}$ norm blow-up in finite 
time.

\section{Conclusion}\label{sec_conc}
In the present paper we have studied numerically the evolution under 
NV dynamics of localized initial data and of perturbed KdV solitons. 

For large values of $ E $ the NV equation behaves qualitatively as the 
KPI equation: sufficiently small KdV solitons appear to be stable under localized 
perturbations. Larger solitons appear to be unstable against the
formation of lump-like structures which resemble the Grinevich-Zakharov traveling wave solutions to NV at $ E > 0 $. The long time behaviour of NV solutions for $E\gg1$ for localized 
initial data is given by lumps and radiation. The results can be 
summarized in the following conjecture: 
\begin{conjecture}
Let $E\gg 1$.\\
- The KdV soliton (\ref{kdvsol}) for small $a$ is stable under NV 
dynamics.\\
- The KdV soliton (\ref{kdvsol}) for large $a$ is unstable under NV dynamics. 
Asymptotically for $t$ large, smaller KdV solitons, lumps and 
radiation will appear. \\
- Localized initial data will develop under NV dynamics for $t$ large 
into radiation and lumps.  
\end{conjecture}

For $E\ll -1$, the NV equation behaves qualitatively as the KPII 
equation: the KdV soliton appears to be stable, and localized initial 
data will be just radiated away:
\begin{conjecture}
Let $E\ll -1$.\\
- The KdV soliton (\ref{kdvsol})  is stable under NV 
dynamics.\\
- Localized initial data will be radiated away to infinity under NV dynamics 
for $t$ large.  
\end{conjecture}

For small values of $|E|$, KdV solitons of sufficiently small 
size are again stable. Larger perturbed solitons will form a blow-up at a finite time. 
Localized initial data of sufficiently small $L^{2}$ norm are 
dispersed away as time goes to infinity.  However, when the initial 
data have a large enough $L^{2}$ norm there appears to be a blow-up 
at finite time. We have not observed the formation of lumps from 
localized initial data. However, based on our knowledge about related 
$2d$ integrable equations, we conjecture that it may be due to the 
fact that the time scales that we consider are not large enough. The 
appearance of oscillations transversal to the direction of 
propagation of the initial pulse at $ E > 0 $ could be a potential 
mechanism for lump formation at larger times.  The results can be 
summarized in the following conjecture (note that numerical studies 
of a blow-up are always challenging and have to be taken with a grain 
of salt):
\begin{conjecture}
Let $|E|\ll 10$.\\
- The KdV soliton (\ref{kdvsol}) for small $a$ is stable under NV 
dynamics.\\
- The KdV soliton (\ref{kdvsol}) for large $a$ is unstable under NV 
dynamics against an $L^{\infty}$ blow-up in finite time. \\
- NV solutions corresponding to localized initial data of sufficiently small 
$L^{2}$ norm are global in time.  Localized initial data of 
sufficiently large $L^{2}$ norm will blow-up in finite time. \\
- A blow-up at time $t^{*}$ is self similar according to the 
scaling (\ref{gKP4}) for $t\sim t^{*}$, 
\begin{equation}
    v \sim \frac{1}{L^{2}}Q\left(\frac{z-z^{*}}{L}\right),\quad L =
    \sqrt{t^{*}-t},
    \label{blowup}
\end{equation}
where $z^{*}$ is the location of the blow-up which appears to be 
finite, and where $Q$ is the lump (\ref{new_lumps}). 

\end{conjecture}

The numerical results indicate that the lumps of NV are stable for 
$E\gg1$ since they appear in the perturbations of the KdV soliton. 
Since this is not the case for $E\sim0$, it can be concluded that 
they are unstable against blow-up in this case. A direct study of 
this important question is, however, not possible with the wanted 
accuracy by the Fourier methods applied in this paper. This is due to 
the fact that rational 
functions because of their slow fall-off to infinity are not well 
approximated by a periodic continuation of their restriction to a 
finite computational domain, in contrast to the Schwartz functions 
studied here. The same is also true for the exact blow-up solutions 
(\ref{tai}) and (\ref{hopper}). For such functions, an approximation 
via polynomials, see \cite{BK} and references therein, would be much 
more efficient. The interesting question in this context would be 
whether the blow-up of the exact solutions which is as the 
 blow-up (\ref{eq:SupCrit_Lt}) is unstable whereas the 
generic one is as observed here given by (\ref{eq:Crit_Lt}). Such a 
situation is known from nonlinear Schr\"odinger equations, see for 
instance the discussion in \cite{sul} and references therein. This 
will be the subject of further research.

\section*{Acknowledgements} The authors are thankful to C. Mu\~noz and 
J.-C. Saut for their interest in the present work and stimulating discussions.


\end{document}